\documentclass[twocolumn,secnumarabic,amssymb, nobibnotes, aps, pra]{revtex4-2}
\usepackage{CJK}
\usepackage[english]{babel}

\usepackage{amsmath}
\usepackage{graphicx}
\usepackage{braket}
\usepackage{color,soul}
\usepackage[colorinlistoftodos]{todonotes}

\usepackage[colorlinks=true, allcolors=blue]{hyperref}

\begin{document}
\begin{CJK*}{UTF8}{SongMT} 

\title{Performance Analysis of Superconductor-constriction-Superconductor Transmon Qubits}
\author{Mingzhao Liu (\CJKfamily{gbsn}{刘铭钊})}
\email{mzliu@bnl.gov}
\affiliation{%
 Center for Functional Nanomaterials, Brookhaven National Laboratory, Upton, NY 11973, USA
}%

\author{Charles T. Black}

\affiliation{%
 Center for Functional Nanomaterials, Brookhaven National Laboratory, Upton, NY 11973, USA
}%

\begin{abstract}
This work presents a computational analysis of a superconducting transmon qubit design, in which the superconductor-insulator-superconductor (SIS) Josephson junction is replaced by a co-planar, superconductor-constriction-superconductor (ScS) nanobridge junction. Within the scope of Ginzburg-Landau theory, we find that the nanobridge ScS transmon has an improved charge dispersion compared to the SIS transmon, with a tradeoff of smaller anharmonicity. These calculations provide a framework for estimating the superconductor material properties and junction dimensions compatible with gigahertz frequency ScS transmon operation.    

\end{abstract}

\maketitle
\end{CJK*}

\section{Introduction}
\begin{figure}
	\centering
	\includegraphics[width=0.48\textwidth]{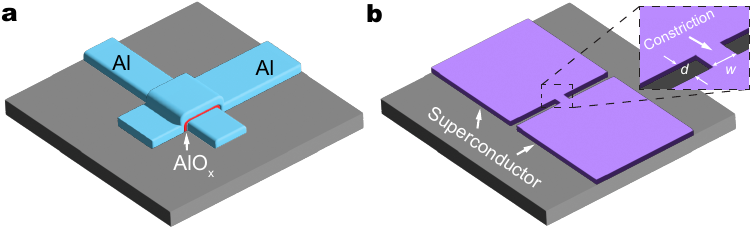}
	\caption[]{(a) Schematic of an Al/AlO$_x$/Al superconductor-insulator-superconductor (SIS) Josephson junction. For clarity, the native oxide covering both Al electrodes is omitted. (b) Schematic of a co-planar superconductor-constriction-superconductor (ScS) Josephson junction, in which two superconducting pads are connected by a nanobridge that has length $d$ and width $w$ (\emph{inset}).}
	\label{SIS_ScS_Scheme}
\end{figure}

The transmon has become an enabling superconducting qubit device architecture, with primary advantages of immunity to charge noise and relatively long coherence lifetimes achieved by designing the device to have Josephson energy far exceeding the charging energy. Similar to other superconducting qubit architectures, the transmon core consists of one or more Josephson junctions (JJs), which are predominantly superconductor-insulator-superconductor tunnel junctions (SIS) -- most often a thin film sandwich structure of aluminum/aluminum oxide/aluminum (Al/AlO$_x$/Al), in which AlO$_x$ is the tunnel barrier (Figure \ref{SIS_ScS_Scheme}a). 

Fabrication of Al/AlO$_x$/Al SIS JJs typically involves physical vapor deposition of the top and bottom Al layers from two different angles relative to the substrate, through a common mask \cite{Dolan1977, Dolan1987}. After depositing the first Al layer, the sample is exposed to a controlled level of oxygen to form the thin AlO$_x$ barrier. The exponential dependence of the JJ critical supercurrent ($I_c$) on tunnel barrier width sets a requirement for tightly controlled oxidation conditions.\cite{Kreikebaum2020, Hertzberg2021} To achieve low device-to-device variation for fabrication at the manufacturing scale, additional considerations must be implemented (e.g., minimizing junction area variations)\cite{Osman2021, Siddiqi2022}. In addition, the temporal drift of $I_c$, commonly referred to as junction aging and attributed to surface chemistry of AlO$_x$ in air, remains an incompletely resolved issuee\cite{Bilmes2021}.  

In a transmon, the SIS JJ is shunted by a large capacitor to minimize the charging energy and thus provide immunity to charge noise. Further, the qubit is coupled to a high-$Q$ microwave resonator for readout. Although high-$Q$ resonators can be fabricated from Al with proper processing, the shunting capacitor and the resonator are more typically fabricated separately from the SIS JJ, using a superconductor with higher $T_c$ and better chemical robustness compared to Al (e.g., niobium ($T_c=9.2$ K)\cite{Premkumar2021}, tantalum ($T_c=4.4$ K)\cite{APMPlace2021}, or titanium nitride ($T_c=5.6$ K)\cite{Pappas2003}). The improved chemical robustness allows the use of post-fabrication wet chemical treatments to remove surface contaminants that contribute to TLS loss.\cite{Crowley2023, McLellan2023} However, some of these types of treatments, particularly those involving strong acids or corrosives,  are more difficult after Al/AlO$_x$/Al junction fabrication, due to the junction's fragile nature \cite{Williams2003}.

In this work we computationally analyze the performance impact of replacing the transmon SIS tunnel junction with a co-planar superconductor-constriction-superconductor (ScS) Josephson junction. A ScS JJ is comprised of two superconductors separated by a nanobridge, or a thin neck of the same superconductor (Figure \ref{SIS_ScS_Scheme}b), with the constriction establishing the superconducting phase difference that enables Josephson behavior. Nanobridge ScS JJs are co-planar and can be fabricated using conventional lithography and metallization. Here, we follow the formalism established by Koch \emph{et al}. in \cite{Koch2007} to determine the electrical properties of ScS transmons, which are shown to be different from SIS transmons, stemming from a different ScS JJ current-phase relationship (CPR) compared to that of a SIS JJ \cite{KO-1, Likharev1979, Golubov2004, Vijay2010}. Comparing the two device architectures, we show that the ScS transmon generally has lower anharmonicity than the SIS transmon, for devices with the same Josephson energy and capacitive energy. However, the smaller anharmonicity is accompanied by a significantly smaller charge dispersion, giving the ScS transmon stronger immunity against charge noise. 

\section{Results and Discussion}

\subsection{General description of ScS transmon anharmonicity}

The Hamiltonian of a transmon can be written as: 
\begin{equation}
\hat{H}=4E_c(\hat{n}-n_g)^2 + E_J(\hat\varphi),
\label{SIS_Hamitonian}
\end{equation}
where $n_g$ is the offset charge and $E_J(\varphi)$ is the potential energy of the junction. The latter is defined by the CPR of the junction, $I_J(\varphi)$, using the following integral:
\begin{equation}
E_J(\varphi) = \int I_JVdt=\int I_J\frac{\Phi_0}{2\pi}\frac{d\varphi}{dt}dt = \int I_J\frac{\Phi_0}{2\pi}d\varphi.
\label{EJ}
\end{equation}
For a SIS junction having a sinusoidal CPR, $I_J(\varphi) = I_c\sin\varphi$, $E_J(\varphi)$ takes a cosine form and the eigenvalues of the Hamiltonian can be solved analytically \cite{Koch2007}. In contrast, the CPR of a nanobridge ScS junction typically distorts away from the sinusoidal form, leading to a modified potential energy, and so in general the eigenvalue problem must be solved numerically.

However, we can estimate the anharmonicity of a ScS transmon using a perturbative approach. The discussion here is limited to a single-valued CPR that is distorted from the sinusoidal form, but retains its $2\pi$-periodicity and odd parity, which set $I_{J}(n\pi)=0$ for all $n\in \mathbb{Z}$.\cite{Likharev1979} Near $\varphi=0$, the CPR can be expressed by a Maclaurin expansion carrying only odd-order terms:
\begin{equation}
    I_{J}(\varphi)=I_0\sum_{n=0}^{\infty} \frac{a_n\varphi^{2n+1}}{(2n+1)!},
    \label{Is_taylor} 
\end{equation}
in which $I_0=I_J'(0)$ and the coefficients $a_{n} = I_J^{(2n+1)}(0)/I_0$ (note this definition sets $a_0\equiv 1$). Note that the CPR of a SIS junction is thus a special case of Eq. \ref{Is_taylor}, with $I_0=I_c$ and $a_{n}=(-1)^n$.

Near $\varphi=0$, the potential energy of a ScS transmon is obtained by integrating the Maclaurin series in Eq. \ref{Is_taylor}, as
\begin{equation}
   E_{J,\textrm{ScS}}(\varphi) = \frac{I_0\Phi_0}{2\pi}\left(\frac{1}{2}\varphi^2+\sum_{n=1}^{\infty} \frac{a_n\varphi^{2n+2}}{(2n+2)!}\right).
\label{EJ_ScS_taylor}
\end{equation}
By comparing Eq. 4 with the potential energy of a SIS transmon,
\begin{equation}
\begin{aligned}
   E_{J,\textrm{SIS}}(\varphi) &= E_{J,\textrm{SIS}}(1-\cos\varphi)\\
   &= E_{J,\textrm{SIS}}\left(\frac{1}{2}\varphi^2+\sum_{n=1}^{\infty} \frac{(-1)^n\varphi^{2n+2}}{(2n+2)!}\right), 
\label{EJ_SIS_taylor}
\end{aligned}
\end{equation}
we may define the Josephson energy of a ScS transmon as $E_{J,\textrm{ScS}}=I_0\Phi_0/2\pi$ and recognize that the anharmonicity, led by a $\varphi^4$ term, is approximately $-a_1$ times that of a SIS transmon.

The leading anharmonic term ($a_1\varphi^4/4!$) in Eq. \ref{EJ_ScS_taylor} can be treated as a first-order perturbation to the quantum harmonic osscillator (QHO), so that the $m$th eigenenergy of a ScS transmon is approximated as:
\begin{equation}
E_{m,\textrm{ScS}}\approx\hbar\omega_p\left(m+\frac{1}{2}\right) + \frac{a_1}{4}\left(2m^2+2m+1\right)E_C,
\label{ScS_anharmonicity}
\end{equation}
in which $\hbar\omega_p=\sqrt{8E_JE_C}$ is the Josephson plasma energy. The transition energy between the $(m-1)$th and $m$th levels is therefore:
\begin{equation}
E_{m-1,m,\textrm{ScS}}\approx\hbar\omega_p + a_1mE_C,
\label{ScS_anharmonicity2}
\end{equation}
which recovers the familiar SIS result with $a_1=-1$. \cite{Koch2007} In general, if $a_1<0$, the ScS transmon will have negative quartic anharmonicity, similar to the SIS transmon\cite{Hioe1975}.

\subsection{Ginzburg-Landau analysis of a ScS junction}

\begin{figure}
	\centering
	\includegraphics[width=0.48\textwidth]{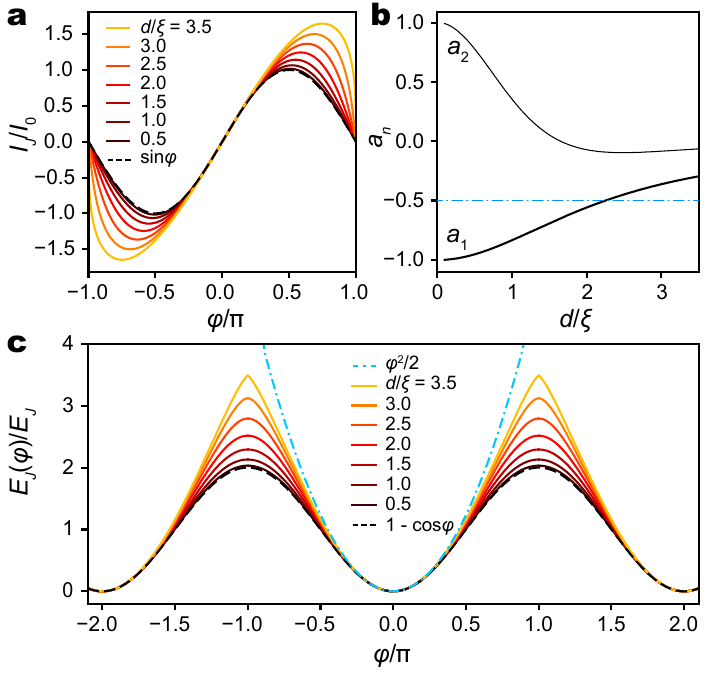}
	\caption[]{(a) The CPR of nanobridge junctions with $0.5\leq d/\xi\leq3.5$, obtained by numerically solving Eqs. \ref{GL_1d} and \ref{GL_CPR}. The sinusoidal curve is shown in dashed black line for comparison. (b) The coefficients $a_1$ and $a_2$ of the Maclaurin series in Eq. \ref{I_single_nb_GL}, as functions of $d/\xi$. (c) The Josephson energy of the nanobridge transmon gradually deviates from the cosine form of SIS junction (dashed black) as $d/\xi$ increases. A harmonic parabola, $\varphi^2/2$, is displayed for reference (dotted cyan).}
	\label{I_single_nb_GL}
\end{figure}

A concrete form of the CPR of a ScS Josephson junction can be derived from an appropriate model of superfluid transport. In this work, we limit our analysis to the approximation based on the Ginzburg-Landau model which, although limited, can provide an estimate of the relationship between the junction nonlinearity and its physical dimensions. Now consider a ScS Josephson junction comprised of a diffusive quasi-one-dimensional nanobridge that is placed in the interval $x\in [-d/2,d/2]$, with length $d$ and width $w$ (Fig. \ref{SIS_ScS_Scheme}b, inset), connecting two large superconducting islands that respectively have uniform order parameters $\Psi=\Psi_0\exp(\pm i\varphi/2)$, where $|\Psi_0|^2=n_s$ is the bulk superfluid density. The two islands thus maintain a phase difference of $\varphi$ across the nanobridge. This neglects any phase drop within the islands, or equivalently, their kinetic inductance. Vijay et al. have noted that the rigid boundary condition is a crude approximation and is valid only when the bridge width $w$ is much shorter than the superconducting coherence length $\xi$.\cite{Vijay2009, Vijay2010, Hasselbach2002} The analysis and following discussion is therefore confined to this regime. With $w\ll\xi$, the spatial variation of the supercurrent density across the nanobridge width can be neglected, so that the order parameter along the nanobridge follows the one dimensional Ginzburg-Landau (GL) equation\cite{Lindelof1981, Likharev1979},

\begin{equation}
    \xi^2\left(\frac{\mathrm{d}}{\mathrm{d}x}-i\frac{2eA}{\hbar}\right)^2\Psi+\left(1-\frac{|\Psi|^2}{|\Psi_0|^2}\right)\Psi=0.
    \label{GL_1d} 
\end{equation}
In the dirty limit, $\xi$ can be approximated by the geometric mean of the Pippard coherence length $\xi_0$ and the electron mean free path $l$, i.e., $\xi\approx\sqrt{\xi_0l}$\cite{TinkhamBook2004}. By neglecting the self-magnetic field and therefore the vector potential $A$, the simplified equation 
 \begin{equation}
    \xi^2\frac{\mathrm{d}^2}{\mathrm{d}x^2}\Psi+\left(1-\frac{|\Psi|^2}{|\Psi_0|^2}\right)\Psi=0
    \label{GL_1d_sim} 
\end{equation}
  can be solved numerically with the boundary conditions of $\Psi(\pm d/2)=\Psi_0\exp(\pm i\varphi/2)$. The solution is then used to compute the CPR, following  
\begin{equation}
    I_{J}(\varphi)=\frac{ie\hbar S}{2m}\left(\Psi^*\frac{\mathrm{d}\Psi}{\mathrm{d}x}-\Psi\frac{\mathrm{d}\Psi^*}{\mathrm{d}x}\right),
    \label{GL_CPR} 
\end{equation}
in which $S$ is the cross-sectional area of the nanobridge. 

As shown in Fig. \ref{I_single_nb_GL}a, the CPR gradually distorts away from the $\sin\varphi$ form of a SIS junction with increasing constriction length ($d/\xi$). The characteristic current $I_0$, as defined in Eq. \ref{Is_taylor}, is given by
\begin{equation}
I_0=\frac{e\hbar S}{md}|\Psi_0|^2,
\label{current_I0}
\end{equation}
to ensure that $a_0=I_J'(0)/I_0=1$. The coefficients $a_1$ and $a_2$ are calculated from the numerical solutions using finite differences, and shown in Fig. \ref{I_single_nb_GL}b for constriction lengths from zero to 3.5$\xi$. Unlike the sinusoidal CPR that has $a_n = (-1)^n$, both $a_1$ and $a_2$ decrease in amplitude toward larger $d/\xi$. According to the plot, 50\% of anharmonicity, as measured by $|a_1|$, is retained when $d/\xi\approx 2.25$. For short constrictions with $d/\xi\ll\sqrt{15}\approx3.87$, Likharev and Yakobson gave an approximate CPR in closed form\cite{Likharev1975c, Likharev1979}, by solving Eq. \ref{GL_1d} through a perturbation approach: 
\begin{equation}
I_{J}(\varphi)=I_0\left[\left(1+\frac{d^2}{15\xi^2}\right)\sin\varphi-\frac{d^2}{30\xi^2}\sin(2\varphi) \right].
\label{LY_CPR} 
\end{equation}
As shown in Appendix \ref{appendix A}, the approximation matches the numerical result for constrictions with length $d/\xi$ up to 1.5, but departs more significantly for longer nanobridges.

The diminishing anharmonicity is more clearly visualized by numerically calculating the potential energy $E_J(\varphi)$ using the integral in Eq. \ref{EJ}. As shown by Fig. \ref{I_single_nb_GL}c, $E_J(\varphi)$ gradually deviates from the cosine curve (dashed black) and approaches the parabola (dotted cyan), as $d/\xi$ increases.

Finally, we note that the GL theory was developed for $T\approx T_c$. Although generally GL remains valid at lower $T$, more precise descriptions of superfluid transport at arbitrary $T$ are available. For example, Kulik and Omelyanchuka have given a solution (KO-1) for short, 1D ScS junctions with $w\ll d$ and $d\ll\xi$ in the dirty limit.\cite{KO-1} As shown in Appendix \ref{appendix B}, the CPR of a KO-1 junction at $T=0$ K has $a_1=-1/2$ and $a_2=0$. By coincidence, this makes the KO-1 CPR numerically very similar as the GL CPR for a nanobridge with length $d/\xi=2.25$. In general, an arbitrary skewness can be introduced into the constriction junction CPR using the form first proposed by Likharev: 
\begin{equation}
I_J=I_c\sin(\varphi - \mathcal{L}I_J),
\label{skewed_sin} 
\end{equation}
in which $\mathcal{L}$ parameterizes the skewness.\cite{Likharev1979} As shown in Appendix \ref{appendix C}, this ``skewed sinusoidal'' CPR behaves nearly identically as the GL CPR, within given range of $\mathcal{L}$. The skewness factor $\mathcal{L}$ has the physical meaning of being the kinetic inductance associated with the constriction junction, \cite{Schussler1993} such that when phase drop in the contact leads (i.e., their kinetic inductance) is non-negligible, we may capture its effect by absorbing the additional kinetic inductance into the $\mathcal{L}$ term. Similarly, we may also view GL $d/\xi$ as a pure parameter, with the value of $d$ approaching the physical length of the nanobridge only in the limit of $w\ll\xi$. For these reasons, we will limit the rest of our discussion to CPRs based on the GL solution, while recognizing that the discussion is generally applicable to CPRs of other forms or derived from a different transport model.

\subsection{Eigenenergies and eigenstates of a ScS transmon} \label{ssec:2.3}

\begin{figure}
	\centering
	\includegraphics[width=.48\textwidth]{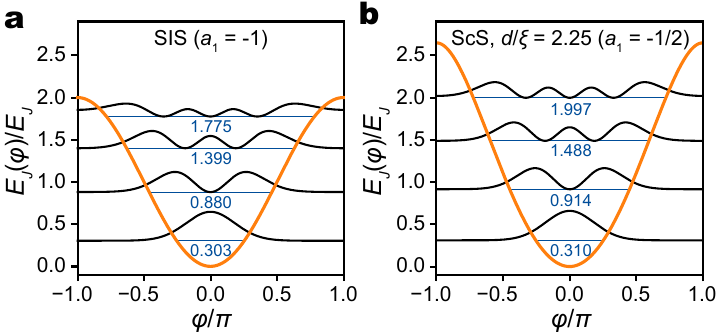}
	\caption[]{The eigenenergies (blue lines and numbers) and the probability densities ($\lVert\Psi\rVert^2$) of the first 4 eigenstates of (a) a SIS transmon and (b) a nanobrige ScS transmon with $d/\xi=2.25$, both with $E_J/E_C=20$ and $n_g=1/2$. The corresponding potential energies, normalized by $E_J$, are plotted in red lines for both transmons.}
\label{Eigenstates}
\end{figure}

\begin{figure*}
	\centering
	\includegraphics[width=\textwidth]{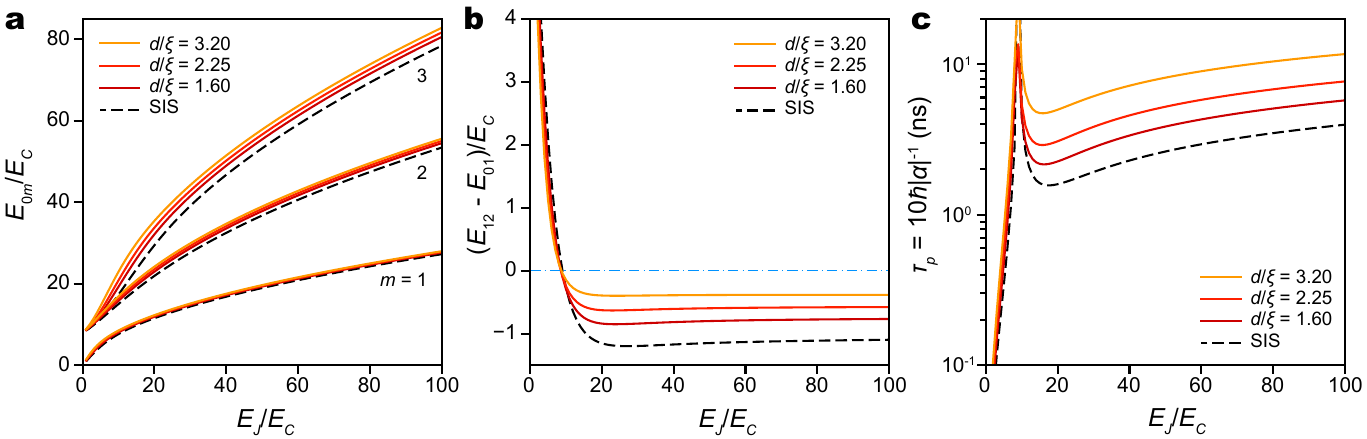}
	\caption[]{(a) Transition energy $E_{0m} = E_m - E_0$ at $n_g = 1/2$, for $m=1$, 2, and 3,  and (b) oscillator anharmonicity ($E_{12}-E_{01}$) at $n_g = 1/2$, as functions of $E_J/E_C$ for SIS transmons (dashed lines) and nanobridge ScS transmons (solid lines). (c) The minimal pulse duration ($\tau_p$) of SIS (dashed line) and ScS transmons (solid line) \emph{vs.} $E_J/E_C$, all operated at $\omega_{01} = 2\pi\times 10$ GHz. In all 3 panels, The nanobridges have lengths $d/\xi$ equal to 1.60 (red), 2.25 (orange), and 3.20 (yellow), with corresponding anharmonicity coefficients of $a_1 = -2/3$, $-1/2$, and $-1/3$, respectively.} 
\label{Anharmonicity}
\end{figure*}

 For a nanobridge ScS transmon with the potential energy illustrated in Fig. \ref{I_single_nb_GL}c, the wave equation can be solved numerically using the finite difference method, in which the Hamiltonian is expressed in a discretized space of phase $\varphi\in[-\pi, \pi)$, with the periodic boundary condition applied to both ends. The validity of our computation is confirmed by comparing a similar numerical solution of the wave equation for a SIS transmon with the analytical solutions presented by Koch \emph{et al} \cite{Koch2007} (see also Appendix \ref{appendix C}). Figure \ref{Eigenstates} compares the first four eigenstates of a SIS transmon and a nanobridge ScS transmon with length $d/\xi=2.25$ and anharmonicity coefficient $a_1=-1/2$, both with $E_J/E_C=20$ and $n_g=1/2$. Although the lower level eigenenergies and eigenfunctions are similar, the differences become more apparent at higher energies. This trend is more clearly observed for the transition energies $E_{0m} = E_m - E_0$ calculated for SIS transmon and the nanobridge ScS transmons at $n_g=1/2$, across a range of $E_J/E_C$ from 1 and 100 (Figure \ref{Anharmonicity}a). The calculation is made for three different nanobridge lengths, $d/\xi$ = 1.60, 2.25, and 3.20, which correspond to anharmonicity coefficients of $a_1 = -2/3$, $-1/2$, and $-1/3$, respectively. In general, ScS transmons with shorter constrictions ($d/\xi$) behave more like SIS transmons, with longer constrictions becoming more QHO-like.
 
 The anharmonicities of ScS transmons, defined as $\alpha = E_{12}-E_{01}$, are plotted against $E_J/E_C$ and compared with SIS transmons (Fig. \ref{Anharmonicity}b), and show diminishing anharmonicity as constriction length $d/\xi$ increases, closely following the perturbation theory result in Eq. \ref{ScS_anharmonicity2}, i.e. $\alpha=a_1E_C$. The smaller anharmonicity of a ScS transmon means that the transitions $E_{01}$ and $E_{12}$ lie closer in energy, so that a longer RF pulse is needed to correctly excite the desired transition $E_{01}$. The minimal pulse duration can be estimated as $\tau_p\approx10\hbar|\alpha|^{-1}$, where the factor of 10 is included for practical reasons such as the requirement for pulse shaping.\cite{Werninghaus2021} The reduced anharmonicity will therefore require longer pulses to maintain the gate fidelity, decreasing gate speed as compared to conventional SIS transmons. As shown in Figure \ref{Anharmonicity}c, despite its lower anharmonicity, $\tau_p$ of the ScS transmon remains less than 10 ns even for $E_J/E_C = 100$, when the qubit operates at 10 GHz. For applications where qubit pulse durations are a few to a few tens of ns, we conjecture that the negative impact of the lower anharmonicity on ScS transmon performance may be tolerable. 

\subsection{Charge dispersion of a ScS transmon}

\begin{figure*}
	\centering
	\includegraphics[width=\textwidth]{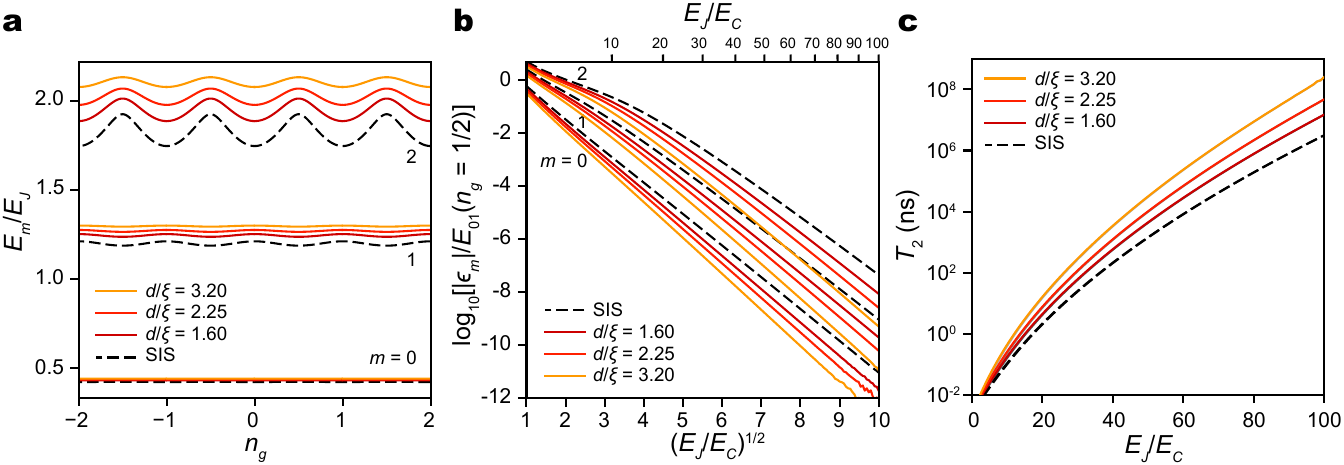}
	\caption[]{(a) The eigenenergies $E_m$ of the lowest 3 eigenstates ($m=0,1,2$) of SIS transmon (dashed line) and nanobridge ScS transmons, as functions of the offset charge $n_g$. All have $E_J/E_C=10$. (b) The charge dispersion $\epsilon_m$ of the lowest 3 eigenstates and the dephasing time $T_2$ of SIS transmons (dashed lines) and nanobridge ScS transmons (solid lines), all operated at $\omega_{01} = 2\pi\times 10$ GHz, as functions of $E_J/E_C$. In all 3 panels, the nanobridges have lengths $d/\xi$ equal to 1.60 (red), 2.25 (orange), and 3.20 (yellow),  corresponding to $a_1 = -2/3$, $-1/2$, and $-1/3$, respectively.}
\label{charge_dispersion}
\end{figure*}
A primary benefit of the transmon architecture is its relative immunity to charge noise, when designed to operate in the regime of $E_J\gg E_C$. In a SIS transmon, the charge dispersion of the $m$th level decreases exponentially with $\sqrt{8E_J/E_C}$, following \cite{Koch2007}:
\begin{equation}
    \begin{aligned}
        \epsilon_m &\equiv E_m(n_g=1/2)-E_m(n_g=0) \\
        &\approx E_C\frac{2^{4m+5}}{(-1)^mm!}\sqrt{\frac{2}{\pi}}\left(\frac{E_J}{2E_C}\right)^{\frac{m}{2}+\frac{3}{4}}\mathrm{e}^{-\sqrt{8E_J/E_C}}.
    \end{aligned}
    \label{SIS_dispersion}
\end{equation}

Intuitively, the charge dispersion is related to the tunneling probability between neighboring potential energy valleys (Fig. \ref{I_single_nb_GL}c), e.g., when $\varphi$ makes a full $2\pi$ rotation \cite{Koch2007}. By this reasoning, we may expect the higher barrier height of a nanobridge ScS transmon to better suppress the tunneling probability and provide lower charge dispersion, compared to a SIS transmon.  

Figure \ref{charge_dispersion}a plots the first three eigenenergies $E_m$ $(m = 0, 1, 2)$ versus the effective offset charge $n_g$ for both SIS (dashed) and nanobridge ScS (solid) transmons, with $E_J/E_C=10$. Clearly, in longer constrictions (i.e., larger $d/\xi$), the ScS transmon eigenenergies are more weakly perturbed by $n_g$. Calculations of the charge dispersion, $\epsilon_m=E_m(n_g=1/2)-E_m(n_g=0)$, across a wide range of $1\leq E_J/E_C\leq 100$ show that suppression of charge dispersion in the ScS transmon becomes more effective for larger $E_J/E_C$ ratios and longer contriction length, $d/\xi$ (Figure \ref{charge_dispersion}b). When $E_J/E_C=100$, the charge dispersion of the first excited state of a ScS transmon with length $d/\xi=2.25$ is over one order of magnitude less than the corresponding SIS transmon. It is noted that the computation for a SIS transmon matches the analytical result very well \cite{Koch2007}, again demonstrating the high numerical precision of our finite difference computation. The computational error only becomes significant as the normalized charge dispersion, $|\epsilon_m|/E_{01}$, approaches $10^{-11}$ and smaller (visible in the lower right corner of Figure \ref{charge_dispersion}b). This is due to the accumulation of floating-point error that eventually shows up when evaluating the vanishing difference between the two eigenenergies at $n_g=0$ and $1/2$. 

In Figure \ref{charge_dispersion}b, the $y-$axis is presented on a logarithmic scale and the $x-$axis is scaled as $\sqrt{E_J/E_C}$, so that all curves take a linear form approaching large $E_J/E_C$ values. For the SIS transmon, the slope matches the expected $\exp(-\sqrt{8E_J/E_C})$ dependence in Eq. \ref{SIS_dispersion}. For the ScS transmon with length $d/\xi=2.25$, the slope is larger, and is best described by: 
\begin{equation}
\epsilon_m\propto\exp\left(-\sqrt{1.17\times8E_J/E_C}\right).
\label{dispersion_slope}
\end{equation} 

The improved charge dispersion makes the ScS transmon less sensitive to charge noise and, in turn, gives it a longer dephasing time $T_2$. For dephasing caused by slow charge fluctuations of large amplitude, Koch \emph{et al}. \cite{Koch2007} has found an upper limit of $T_2$ given by:
\begin{equation}
T_2\approx \frac{4\hbar}{\mathrm{e}^2\pi|\epsilon_1|}.
\label{T2_estimate}
\end{equation}
Using this relation, we compare $T_2$ for both SIS and ScS transmons for $E_J/E_C$ between 1 and 100 (Figure \ref{charge_dispersion}c). The ScS transmon improves $T_2$ across the entire range of $E_J/E_C$ and especially at higher ratios. At $E_J/E_C=100$, the SIS transmon has a $T_2$ ceiling of about 3 ms, compared to about 15 ms for the ScS transmon with $d/\xi=1.60$ ($a_1=-2/3$), a 5-fold increase. More significantly improvement is achieved for longer constrictions (i.e., larger $d/\xi$). At present, because the $T_1$ lifetime of SIS transmon qubits is still beyond 1 ms and not limited by the charge noise, this advantage of the ScS transmon architecture will yield little performance benefit. However, because we expect the lifetimes of superconducting qubits to continue improving (\emph{Schoelkopf's Law}) \cite{Kjaergaard2020}, there may be a point when charge noise dephasing becomes a bottleneck, for which the ScS transmon architecture can offer effective mitigation despite the tradeoff of slower gate speed.

\begin{figure*}
	\centering
	\includegraphics[width=\textwidth]{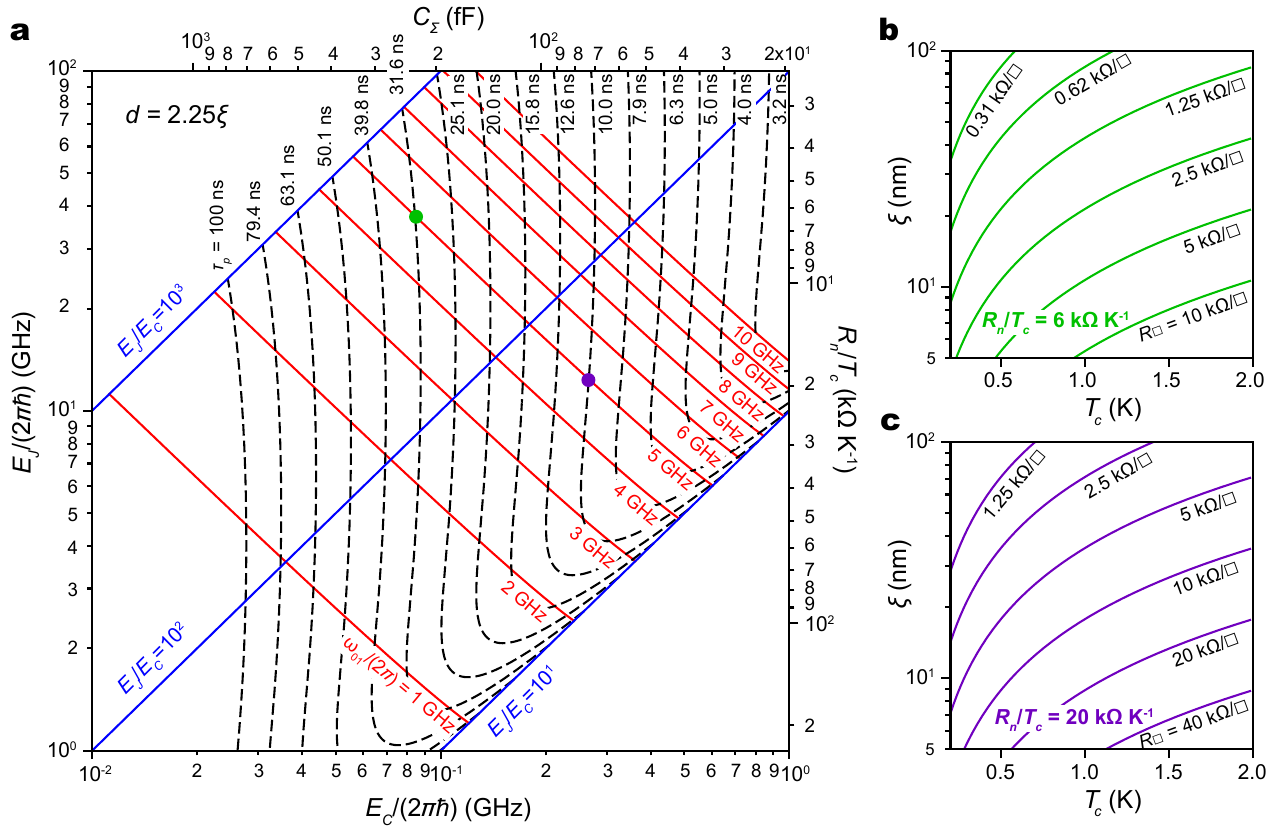}
\caption[]{(a) A graphical guide for designing a nanobridge ScS transmon with required $E_J$ and $E_C$ to match desired transmon frequency $\omega_{01}$ and minimum pulse duration $\tau_p$. The nanobridge length is fixed at $2.25\xi$. The red lines are contours lines for transmon frequencies set at values between 1 and 10 GHz. The dashed black lines are contours lines for $\tau_p$ set at a few values between 3.2 and 100 ns. The blue lines are contours lines for $E_J/E_C$ ratios set at 10, 100, and 1000. A second $x$-axis that is parallel to $E_C$ is presented for $C_\Sigma$, following $C_\Sigma = e^2/2E_C$. Simlarly, a second $y$-axis that is parallel to $E_J$ is presented for $R_n/T_c$, following $R_n/T_c=1.76k_B\Phi_0/(2eE_J)$. (b, c) The normal state sheet resistances $R_\square$ required to match the $R_n/T_c$ values at the (b) green spot (6 $\mathrm{ k\Omega\cdot K^{-1}}$) and the (c) purple spot (20 $\mathrm{ k\Omega\cdot K^{-1}}$), for materials with different values of $T_c$ and $\xi$. The nanobridge width is fixed at 20 nm.}  
\label{design_parameters}
\end{figure*}

\subsection{ScS transmon design parameters}

The operational behavior of a ScS transmon is determined by its $E_J$, $E_C$, and anharmonicity $\alpha$, which define the operating frequency $\omega_{01}$, the relative immunity to charge noise ($\epsilon_1$), and the minimum excitation pulse duration ($\tau_p$). These three quantities are not independent. We can visualize this interdependence with three sets of contour lines plotted in the plane of $E_J$ versus $E_C$ (Figure \ref{design_parameters}a), in which we set the anharmonicity to $\alpha = -E_C/2$, corresponding to a nanobridge ScS with length $d/\xi=2.25$.  These contours represent: (1) a transmon operating frequency ($\omega_{01}/(2\pi)$) between 1 and 10 GHz (set of red, descending diagonal lines), (2) ratios of $E_J/E_C$ from 10, 100, and 1000 (set of blue, ascending diagonal lines), and (3) the minimum excitation pulse duration $\tau_p$ between 3.2 and 100 ns (set of dashed, predominantly vertical lines). Selecting two of these defines the third one. For example, a ScS transmon designed to operate at $\omega_{01}/(2\pi)=5$ GHz and with an excitation  pulse of $\tau_p=32$ ns (green dot in Figure \ref{design_parameters}) must have a $E_J/E_C$ ratio of about 400. Instead, a shorter excitation pulse of $\tau_p=10$ ns (purple dot in Figure \ref{design_parameters}) requires a tradeoff of smaller $E_J/E_C\approx40$, and thus less immunity against charge noise.

Importantly, $E_J$ and $E_C$ of a ScS transmon are set by the physical device dimensions and fundamental properties of the materials composing it. $E_J$ is given by $I_0\Phi_0/2\pi$, in which the characteristic current $I_0$ may be approximated by $\pi\Delta/eR_n$ in the dirty limit, where $\Delta$ is the superconducting energy gap and $R_n$ is the normal state resistance of the junction. For a BCS superconductor where $\Delta = 1.76k_BT_c$, we can express $E_J$ in terms of the material properties $R_n/T_c=1.76k_B\Phi_0/(2eE_{J,\textrm{ScS}})$, which is shown as the second (right) $y$-axis in Figure \ref{design_parameters}a. Similarly, because $E_C$ is set by the total capacitance ($C_\Sigma = e^2/2E_{C,\textrm{ScS}}$) which depends on device geometry and dielectric properties, we can express $E_C$ as a capacitance, shown as a second (top) $x$-axis in Figure \ref{design_parameters}a.  

Returning to the example, we can now see from Fig. \ref{design_parameters}a that designing a nanobridge ScS transmon with $\omega_{01}/(2\pi)=5$ GHz, $\tau_p=32$ ns, and $E_J/E_C$ ratio of about 400 (green dot) requires a junction with $R_n/T_c\approx6$ $\mathrm{ k\Omega\cdot K^{-1}}$ and capacitor with $C_\Sigma\approx 250$ fF. If one instead desires the shorter excitation pulse time of $\tau_p=10$ ns (purple dot), the constriction must have $R_n/T_c\approx20$ $\mathrm{k\Omega\cdot K^{-1}}$ and $C_\Sigma\approx 75$ fF. Because $R_n$ is related to the constriction sheet resistance through $R_n = R_\square d/w$, and $d$ is fixed at $2.25\xi$ in this example, the following constraint applies to the values of material parameters ($R_\square$, $T_c$, $\xi$) and nanobridge width $w$:
\begin{equation}
\frac{R_n}{T_c}=\frac{2.25R_\square\xi}{T_cw},
\label{parameter_constraint_1}
\end{equation}
to design a nanobridge to any $R_n/T_c$ value in Fig. \ref{design_parameters}a. 

Equation \ref{parameter_constraint_1} shows that to achieve the $R_n/T_c$ values of a few $\mathrm{k\Omega\cdot K^{-1}}$ required for 5 GHz transmon operation, $R_\square$ and $\xi$ should be large and $T_c$ and $w$ small. The nanobridge width $w$ is limited by available nanofabrication techniques. Setting $w = 20$ nm, the constraints on $R_\square$, $T_c$, and $\xi$ applied by Eq. \ref{parameter_constraint_1} are illustrated in Figures \ref{design_parameters}b and \ref{design_parameters}c for $R_n/T_c$ values specified by the green and purple dots in Figure \ref{design_parameters}a, respectively. The requirement for large $R_\square$ is eased with lower $T_c$. However, $T_c$ must be high enough to minimized quasiparticle loss at the working temperature of transmon (typically 20 mK), so we set the lowest $T_c$ to 0.2 K in this example.

These plots show that, in general, $R_\square$ must be $\sim1-10$ $\mathrm{k\Omega/\square}$, which is difficult to achieve in conventional metallic superconductors, but is within the range of high kinetic inductance (KI) superconductors, such as granular aluminum\cite{Rotzinger2017, Maleeva2018} or amorphous niobium-silicon alloys ($a$-Nb$_x$Si$_{1-x}$)\cite{Dumoulin2008,Dumoulin2013, Webster2013}. The latter material is also promising because of its relatively long coherence length ($30-50$ nm) and suitable $T_c$ ($0.3-1$ K)\cite{Dumoulin2008, Dumoulin2013}. 

Lastly, we note that the requirement for large $R_n$ can be relaxed for transmons operated at higher frequency approaching the range of millimeter waves. For example, by extrapolating the plot in Figure \ref{design_parameters}a, we can estimate that a transmon operated at $\omega_{01}=2\pi\times 20$ GHz with $E_J/E_C = 500$ requires $\tau_p = 10$ ns and $R_n/T_c$ only about 1 $\mathrm{k\Omega\cdot K^{-1}}$, which can be achieved with a much wider selection of materials. In general, because ScS junctions can support large critical current densities, they have advantages for higher frequency operations that requires large $E_J$ (i.e., $I_c$).

\section{Conclusion}

In summary, we have shown through computation that a short ScS Josephson junction can be used as a substitute for the SIS tunnel junction in a transmon qubit. In the transmon regime ($E_J\gg E_C$), a ScS transmon has smaller anharmonicity than a SIS transmon, but appreciably lower charge dispersion that provides a significantly higher $T_2$ ceiling. Using this analysis, we estimate that high performance ScS transmons can be achieved with narrow constrictions (i.e., $w\ll\xi$) having a normal state resistance of a few kOhms, which can be made from a thin nanobridge formed in low $T_c$, high KI superconductors using conventional, high-resolution nanofabrication techniques. The ScS transmon design may allow all components, including constriction junction, capacitor, and resonator, to be fabricated in a single lithography step. This is a simplification compared to conventional SIS transmon fabrication, and may provide an architecture amenable to device post-processing, cleaning, and encapsulation.

\begin{acknowledgments}
This material is based upon work supported by the U.S. Department of Energy, Office of Science, National Quantum Information Science Research Centers, Co-design Center for Quantum Advantage (C$^2$QA) under contract number DE-SC0012704. This research used computational resources of the Center for Functional Nanomaterials (CFN), which is a U.S. Department of Energy Office of Science User Facility, at Brookhaven National Laboratory under Contract No. DE-SC0012704. Helpful discussion with Prof. K. K. Likharev is greatly acknowledged.
\end{acknowledgments}

\appendix

\section{Nanobridge ScS CPR according to Likharev-Yakobson approximation}
\label{appendix A}

\begin{figure}[b]
	\centering
	\includegraphics[width=.48\textwidth]{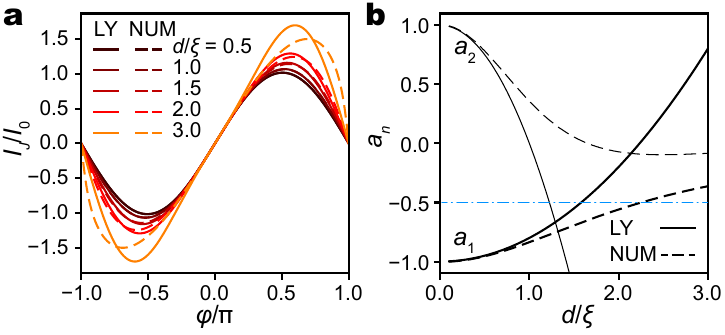}
	\caption[]{(a) The CPRs of nanobridge junctions according to the closed form approximation by Likharev and Yakobson (LY, solid lines), compared with the numerical results (NUM, dashed lines). (b) The Maclaurin coefficients $a_1$ and $a_2$, as functions of $d/\xi$, for LY solution and the numerical solution.}
\label{LY_CPR_EJ}
\end{figure}

Likharev and Yakobson (LY) gave an approximated solution to  Eq. \ref{GL_1d_sim}\cite{Likharev1975c, Likharev1979}, using a perturbative approach. When the nanobridge and the superconducting islands have the same $T_c$, the corresponding CPR is written,
\begin{equation}
\begin{aligned}
I_{J}(\varphi)=\frac{e\hbar S}{md}|\Psi_0|^2&\Bigg[\left(1+\frac{d^2}{15\xi^2}\right)\sin\varphi\\
&-\frac{d^2}{30\xi^2}\sin(2\varphi) \Bigg].
\end{aligned}
\label{LY_CPR_app} 
\end{equation}
Maclaurin expansion according to Eq. \ref{Is_taylor} produces $I_0=\frac{e\hbar S}{md}|\Psi_0|^2$, $a_1=I_J^{(3)}(0)/I_0=-1 + d^2/(5\xi^2)$, and $a_2=I_J^{(5)}(0)/I_0=1-d^2/\xi^2$. For a few selected values of $d/\xi$, the LY CPRs are compared with our numerical solutions in Fig. \ref{LY_CPR_EJ}a, showing that the two solutions match each other up to $d/\xi=1.5$. A comparison of their Maclaurin coefficients further reveals that the LY solution approaches the numerical solution as $d/\xi\rightarrow 0$, but diverges quickly as $d/\xi$ exceeds 1 (Fig. \ref{LY_CPR_EJ}b).

\section{ScS CPR according to KO-1 model}
\label{appendix B}
 
A more precise description of superfluid transport at arbitrary $T$ is given by the Eilenberger equations, which further reduces to the Usadel equation in the dirty limit. For a short ScS junction with  $w\ll d$ and $d\ll\xi$, a solution was given by Kulik and Omelyanchuk (KO-1). At $T=0$ K, the KO-1 CPR writes:
\begin{equation}
I_J(\varphi)=\frac{\pi\Delta_0}{eR_n}\cos\frac{\varphi}{2}\tanh^{-1}\left(\sin\frac{\varphi}{2}\right),
\label{singlebridgeCPR}
\end{equation}
in which $\Delta_0$ is the superconducting energy gap at 0 K.\cite{KO-1, Golubov2004} Integrating the CPR by Eq. \ref{EJ} produces the potential energy of KO-1 ScS junction at 0 K:
\begin{equation}
\begin{aligned}
E_{J,\textrm{ScS}}(\varphi) &= \frac{\Delta\Phi_0}{2eR_n}\Big[\ln\left(\cos^2\frac{\varphi}{2}\right)\\
&+2\sin\frac{\varphi}{2}\tanh^{-1}\left(\sin\frac{\varphi}{2}\right)\Big].
\end{aligned}
\label{EJ_singleNB}
\end{equation}

\begin{figure}
	\centering
	\includegraphics[width=.48\textwidth]{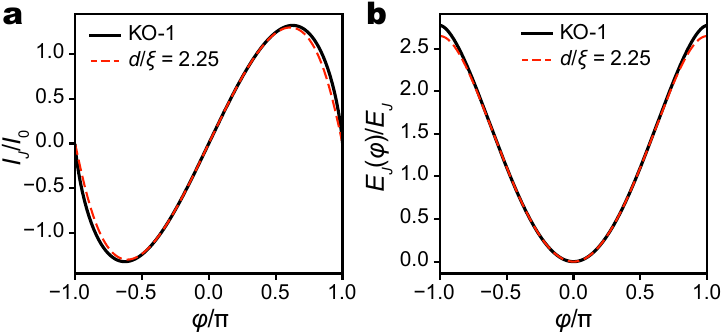}
	\caption[]{(a) The CPR and (b) the potential energy $E_J(\phi)$ of a KO-1 ScS junction at $T=0$ K (black solid lines) are compared with the corresponding ones of a nanobridge ScS with $d/\xi=2.25$ at the GL regime (red dashed lines).}
\label{KO1_CPR_EJ}
\end{figure}

According to the convention set in Eq. \ref{Is_taylor}, Maclaurin expansion of the CPR in Eq. \ref{singlebridgeCPR} produces $I_0=I_J'(0)=\pi\Delta_0/(2eR_n)$, $a_1=I_J^{(3)}(0)/I_0=-0.5$, and $a_2=I_J^{(5)}(0)/I_0=0$. This indicates that, numerically, a short ScS junction in the KO-1 limit will behave very similarly as a nanobridge junction with $d/\xi=2.25$ at the GL limit ($a_1=-0.5$, $a_2=-0.09$), with anharmonicity about \emph{one half} as a SIS junction (Fig. \ref{KO1_CPR_EJ}).

\section{CPR in skewed sinusoidal form}
\label{appendix C}

\begin{figure}
	\centering
	\includegraphics[width=.48\textwidth]{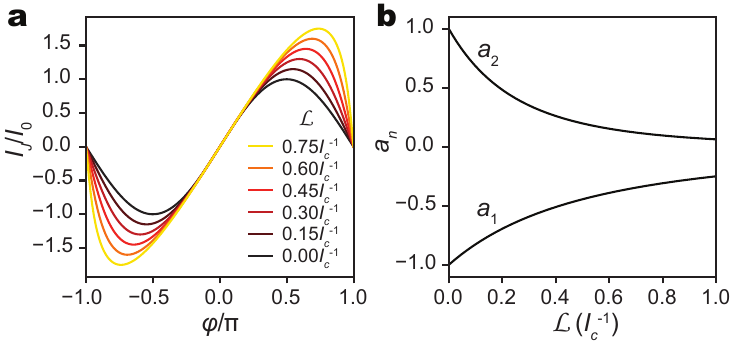}
	\caption[]{(a) The ``skewed sinusoidal'' CPRs according to the numerical form in Eq. \ref{skewed_sin}, with different values of skewness factor $\mathcal{L}$, in the unit of $I_c^{-1}$. (b) The Maclaurin coefficients $a_1$ and $a_2$, as functions of $\mathcal{L}$, according to Eq. \ref{skew_sin_coefficients}.}
\label{skews_sin_CPR_fig}
\end{figure}

The ``skewed sinusoidal'' CPR described by Eq. \ref{skewed_sin}, $I_J=I_c\sin(\varphi - \mathcal{L}I_J)$, does not have a closed-form expression for $I_J$. To obtain its Maclaurin expansion according to Eq. \ref{Is_taylor}, we note that at $\varphi=0$, its first derivative can be expressed as $\mathrm{d}I_J=I_c\mathrm{d}(\varphi-\mathcal{L}I_J)$, which can be rearranged to $\mathrm{d}\varphi=(1+\mathcal{L}I_c)\mathrm{d}(\varphi-\mathcal{L}I_J)$. Now with the second equation relating $\mathrm{d}\varphi$ to the differential of the argument $\varphi-\mathcal{L}I_J$, we may readily find the derivatives of $I_J$ at $\varphi=0$, as
\begin{equation}
I_J^{(2n+1)}(0)=\frac{(-1)^nI_c}{(1+\mathcal{L}I_c)^{2n+1}}.
\label{skew_sin_derivatives}
\end{equation}
Therefore, coefficients for the Maclaurin expansion in Eq. \ref{Is_taylor} are given by
\begin{equation}
\begin{aligned}
I_0 &= I_J'(0)=\frac{I_c}{1+\mathcal{L}I_c},\\
a_n&=I_J^{(2n+1)}(0)/I_0=\frac{(-1)^n}{(1+\mathcal{L}I_c)^{2n}}.
\end{aligned}
\label{skew_sin_coefficients}
\end{equation}
The magnitude of coefficients $a_n$ would therefore decrease with increasing $\mathcal{L}$ (Fig. \ref{skews_sin_CPR_fig}b), following a similar trend as the GL CPR toward longer bridge lengths (Fig. \ref{LY_CPR_EJ}b, dashed lines). As shown in Fig. \ref{skews_sin_CPR_fig}a, the ``skewed sinusoidal'' CPRs, normalized by $I_0$, appear similar to the GL CPRs shown in Fig. \ref{I_single_nb_GL}a.

\section{Finite difference solutions of transmon wave equations}
\label{appendix D}

\begin{figure*}
	\centering
	\includegraphics[width=\textwidth]{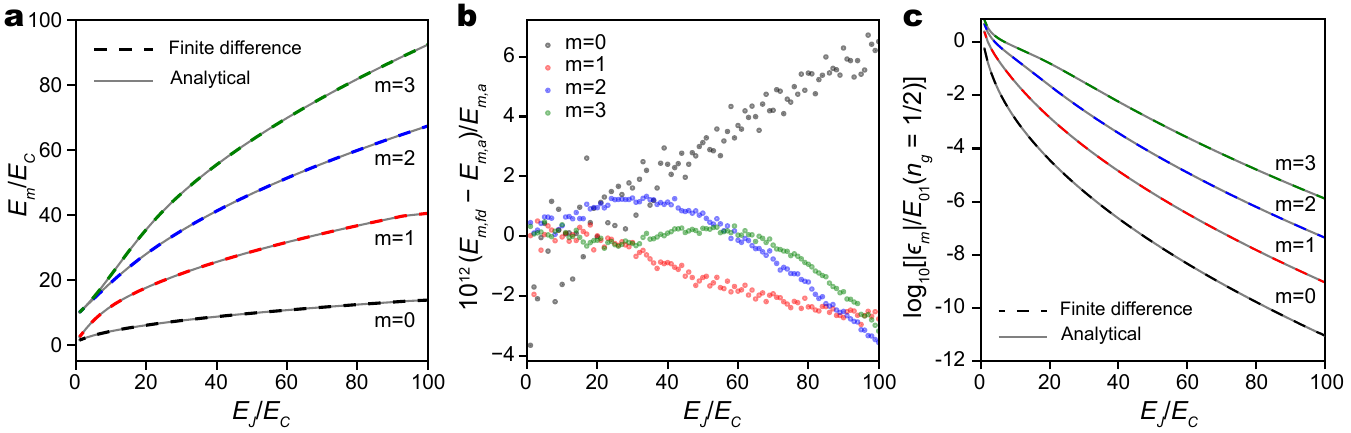}
	\caption[]{(a) The first four eigenenergies of a SIS transmon, plotted versus $E_J/E_C$, for $n_g=1/2$, calculated using the analytical solutions (gray solid lines) and the finite difference method (dashed lines), as functions of $E_J/E_C$. (b) Relative errors of the finite difference solutions, calculated as $(E_{m,\mathrm{fd}}-E_{m,\mathrm{a}})/E_{m,\mathrm{a}}$, in which $E_{m,\mathrm{fd}}$ and $E_{m,\mathrm{a}}$ are the $m$th eigenenergies given by the finite difference method and the analytical solution, respectively. (c) The charge dispersion $\epsilon_m$ of the lowest four eigenstates of a SIS transmon (dashed lines), calculated by using the analytical solutions (gray solid lines) and the finite difference method (dashed lines), as functions of $E_J/E_C$.}
\label{Eigenenergy_verfication}
\end{figure*}

In the phase basis, the kinetic energies of SIS and ScS transmons are both given by
\begin{equation}
\begin{aligned}
T&=4E_C(\hat{n}-n_g)^2 \\&= 4E_C\left(-\frac{\mathrm{d}^2}{\mathrm{d}\varphi^2}+2in_g\frac{\mathrm{d}}{\mathrm{d}\varphi}+n_g^2\right),
\label{kinetic_energy_phase_basis}
\end{aligned}
\end{equation}
by recognizing that $\hat{n}=-i\mathrm{d}/\mathrm{d}\varphi$. The derivative terms can be approximated by finite differences. By discretizing $\varphi$ over the interval $[-\pi,\pi)$ into an arithmetic sequence $\{\varphi_k=-\pi+k\delta\}$, where $k=0,1,\dots,N-1$ and $\delta=2\pi/N$, an analytic function $\psi$ defined over the interval is discretized as a series $\{\psi_k=\psi(\varphi_k)\}$. The first and second derivatives of $\psi$ are approximated by finite differences to the sixth order accuracy of $\delta$ \cite{Fornberg1988}, as: 
\begin{equation}
\begin{aligned}
\psi^{(1)}(\varphi_k)\approx\delta^{-1}\Big(-\frac{\psi_{k-3}}{60}+\frac{3\psi_{k-2}}{20}&-\frac{3\psi_{k-1}}{4}\\
+\frac{3\psi_{k+1}}{4}-\frac{3\psi_{k+2}}{20}&+\frac{\psi_{k+3}}{60}\Big),
\end{aligned}
\label{finite_difference_1}
\end{equation}
and
\begin{equation}
\begin{aligned}
\psi^{(2)}(\varphi_k)\approx\delta^{-2}\Big(\frac{\psi_{k-3}}{90}-\frac{3\psi_{k-2}}{20}&+\frac{3\psi_{k-1}}{2}-\frac{49\psi_k}{18}\\
+\frac{3\psi_{k+1}}{2}-\frac{3\psi_{k+2}}{20}&+\frac{\psi_{k+3}}{90}\Big),
\end{aligned}
\label{finite_difference_2}
\end{equation}

In the discretized space of $\varphi$, the Hamiltonian is expressed by a $N\times N$ matrix $\mathbf{H} = \{H_{kl}\}$, $(k,l=0,1,\dots,N-1)$, which has the following non-zero matrix elements:
\begin{equation}
H_{kl} =
\begin{cases}
\displaystyle{\frac{E_C}{9\delta^2}(98+36n_g^2\delta^2) + E_J(\varphi_k)}, \\
\qquad\qquad\qquad(l=k)\\
\displaystyle{-\frac{6E_C}{\delta^2}\left(1\mp in_g\delta\right)},\\
\qquad\qquad\qquad(l=(k\pm1)\mod N)\\
\displaystyle{\frac{3E_C}{5\delta^2}\left(1\mp 2in_g\delta\right)},\\
\qquad\qquad\qquad(l=(k\pm2)\mod N)\\
\displaystyle{-\frac{2E_C}{45\delta^2}\left(1\mp 3in_g\delta\right)},\\
\qquad\qquad\qquad(l=(k\pm3)\mod N)
\end{cases}
\label{matrix_elements}
\end{equation}
while all the other elements are 0. The modular arithmetic gives a cyclic symmetry to the off-diagonal matrix elements, which enforces the periodic boundary conditions at $\varphi=\pm\pi$. 

The first few eigenvalues and eigenvectors of $\mathbf{H}$ are obtained by using the SciPy \cite{2020SciPy-NMeth} sparse matrix eigensolver (\texttt{scipy.sparse.linalg.eigs}). The validity of the finite difference method is verified by comparing the numerical results for SIS transmon with the analytical Mathieu function solutions. For the first four eigenenergies of SIS transmons with $1\leq E_J/E_C\leq 100$ and $n_g=1/2$, the two methods give virtually identical results (Fig. \ref{Eigenenergy_verfication}a). The relative differences between eigenenergies obtained from the two methods are less than $10^{-11}$, as shown in Fig. \ref{Eigenenergy_verfication}b. The computation results on charge dispersion $\epsilon_m$ are compared in Fig. \ref{Eigenenergy_verfication}c, showing that the two methods match each other very well, further demonstrating the accuracy of the finite difference method.


\begin{thebibliography}{34}%
\makeatletter
\providecommand \@ifxundefined [1]{%
 \@ifx{#1\undefined}
}%
\providecommand \@ifnum [1]{%
 \ifnum #1\expandafter \@firstoftwo
 \else \expandafter \@secondoftwo
 \fi
}%
\providecommand \@ifx [1]{%
 \ifx #1\expandafter \@firstoftwo
 \else \expandafter \@secondoftwo
 \fi
}%
\providecommand \natexlab [1]{#1}%
\providecommand \enquote  [1]{``#1''}%
\providecommand \bibnamefont  [1]{#1}%
\providecommand \bibfnamefont [1]{#1}%
\providecommand \citenamefont [1]{#1}%
\providecommand \href@noop [0]{\@secondoftwo}%
\providecommand \href [0]{\begingroup \@sanitize@url \@href}%
\providecommand \@href[1]{\@@startlink{#1}\@@href}%
\providecommand \@@href[1]{\endgroup#1\@@endlink}%
\providecommand \@sanitize@url [0]{\catcode `\\12\catcode `\$12\catcode
  `\&12\catcode `\#12\catcode `\^12\catcode `\_12\catcode `\%12\relax}%
\providecommand \@@startlink[1]{}%
\providecommand \@@endlink[0]{}%
\providecommand \url  [0]{\begingroup\@sanitize@url \@url }%
\providecommand \@url [1]{\endgroup\@href {#1}{\urlprefix }}%
\providecommand \urlprefix  [0]{URL }%
\providecommand \Eprint [0]{\href }%
\providecommand \doibase [0]{https://doi.org/}%
\providecommand \selectlanguage [0]{\@gobble}%
\providecommand \bibinfo  [0]{\@secondoftwo}%
\providecommand \bibfield  [0]{\@secondoftwo}%
\providecommand \translation [1]{[#1]}%
\providecommand \BibitemOpen [0]{}%
\providecommand \bibitemStop [0]{}%
\providecommand \bibitemNoStop [0]{.\EOS\space}%
\providecommand \EOS [0]{\spacefactor3000\relax}%
\providecommand \BibitemShut  [1]{\csname bibitem#1\endcsname}%
\let\auto@bib@innerbib\@empty
\bibitem [{\citenamefont {Dolan}(1977)}]{Dolan1977}%
  \BibitemOpen
  \bibfield  {author} {\bibinfo {author} {\bibfnamefont {G.~J.}\ \bibnamefont
  {Dolan}},\ }\bibfield  {title} {\bibinfo {title} {Offset masks for lift-off
  photoprocessing},\ }\href {https://doi.org/10.1063/1.89690} {\bibfield
  {journal} {\bibinfo  {journal} {Appl. Phys. Lett.}\ }\textbf {\bibinfo
  {volume} {31}},\ \bibinfo {pages} {337} (\bibinfo {year} {1977})}\BibitemShut
  {NoStop}%
\bibitem [{\citenamefont {Fulton}\ and\ \citenamefont
  {Dolan}(1987)}]{Dolan1987}%
  \BibitemOpen
  \bibfield  {author} {\bibinfo {author} {\bibfnamefont {T.~A.}\ \bibnamefont
  {Fulton}}\ and\ \bibinfo {author} {\bibfnamefont {G.~J.}\ \bibnamefont
  {Dolan}},\ }\bibfield  {title} {\bibinfo {title} {Observation of
  single-electron charging effects in small tunnel junctions},\ }\href
  {https://doi.org/10.1103/PhysRevLett.59.109} {\bibfield  {journal} {\bibinfo
  {journal} {Phys. Rev. Lett.}\ }\textbf {\bibinfo {volume} {59}},\ \bibinfo
  {pages} {109} (\bibinfo {year} {1987})}\BibitemShut {NoStop}%
\bibitem [{\citenamefont {Kreikebaum}\ \emph {et~al.}(2020)\citenamefont
  {Kreikebaum}, \citenamefont {O'Brien}, \citenamefont {Morvan},\ and\
  \citenamefont {Siddiqi}}]{Kreikebaum2020}%
  \BibitemOpen
  \bibfield  {author} {\bibinfo {author} {\bibfnamefont {J.~M.}\ \bibnamefont
  {Kreikebaum}}, \bibinfo {author} {\bibfnamefont {K.~P.}\ \bibnamefont
  {O'Brien}}, \bibinfo {author} {\bibfnamefont {A.}~\bibnamefont {Morvan}},\
  and\ \bibinfo {author} {\bibfnamefont {I.}~\bibnamefont {Siddiqi}},\
  }\bibfield  {title} {\bibinfo {title} {Improving wafer-scale {Josephson}
  junction resistance variation in superconducting quantum coherent circuits},\
  }\href {https://doi.org/10.1088/1361-6668/ab8617} {\bibfield  {journal}
  {\bibinfo  {journal} {Supercond. Sci. Technol.}\ }\textbf {\bibinfo {volume}
  {33}},\ \bibinfo {pages} {06LT02} (\bibinfo {year} {2020})}\BibitemShut
  {NoStop}%
\bibitem [{\citenamefont {Hertzberg}\ \emph {et~al.}(2021)\citenamefont
  {Hertzberg}, \citenamefont {Zhang}, \citenamefont {Rosenblatt}, \citenamefont
  {Magesan}, \citenamefont {Smolin}, \citenamefont {Yau}, \citenamefont
  {Adiga}, \citenamefont {Sandberg}, \citenamefont {Brink}, \citenamefont
  {Chow},\ and\ \citenamefont {Orcutt}}]{Hertzberg2021}%
  \BibitemOpen
  \bibfield  {author} {\bibinfo {author} {\bibfnamefont {J.~B.}\ \bibnamefont
  {Hertzberg}}, \bibinfo {author} {\bibfnamefont {E.~J.}\ \bibnamefont
  {Zhang}}, \bibinfo {author} {\bibfnamefont {S.}~\bibnamefont {Rosenblatt}},
  \bibinfo {author} {\bibfnamefont {E.}~\bibnamefont {Magesan}}, \bibinfo
  {author} {\bibfnamefont {J.~A.}\ \bibnamefont {Smolin}}, \bibinfo {author}
  {\bibfnamefont {J.-B.}\ \bibnamefont {Yau}}, \bibinfo {author} {\bibfnamefont
  {V.~P.}\ \bibnamefont {Adiga}}, \bibinfo {author} {\bibfnamefont
  {M.}~\bibnamefont {Sandberg}}, \bibinfo {author} {\bibfnamefont
  {M.}~\bibnamefont {Brink}}, \bibinfo {author} {\bibfnamefont {J.~M.}\
  \bibnamefont {Chow}},\ and\ \bibinfo {author} {\bibfnamefont {J.~S.}\
  \bibnamefont {Orcutt}},\ }\bibfield  {title} {\bibinfo {title}
  {Laser-annealing {Josephson} junctions for yielding scaled-up superconducting
  quantum processors},\ }\href {https://doi.org/10.1038/s41534-021-00464-5}
  {\bibfield  {journal} {\bibinfo  {journal} {npj Quantum Inf.}\ }\textbf
  {\bibinfo {volume} {7}},\ \bibinfo {pages} {129} (\bibinfo {year}
  {2021})}\BibitemShut {NoStop}%
\bibitem [{\citenamefont {Osman}\ \emph {et~al.}(2021)\citenamefont {Osman},
  \citenamefont {Simon}, \citenamefont {Bengtsson}, \citenamefont {Kosen},
  \citenamefont {Krantz}, \citenamefont {P.~Lozano}, \citenamefont
  {Scigliuzzo}, \citenamefont {Delsing}, \citenamefont {Bylander},\ and\
  \citenamefont {Fadavi~Roudsari}}]{Osman2021}%
  \BibitemOpen
  \bibfield  {author} {\bibinfo {author} {\bibfnamefont {A.}~\bibnamefont
  {Osman}}, \bibinfo {author} {\bibfnamefont {J.}~\bibnamefont {Simon}},
  \bibinfo {author} {\bibfnamefont {A.}~\bibnamefont {Bengtsson}}, \bibinfo
  {author} {\bibfnamefont {S.}~\bibnamefont {Kosen}}, \bibinfo {author}
  {\bibfnamefont {P.}~\bibnamefont {Krantz}}, \bibinfo {author} {\bibfnamefont
  {D.}~\bibnamefont {P.~Lozano}}, \bibinfo {author} {\bibfnamefont
  {M.}~\bibnamefont {Scigliuzzo}}, \bibinfo {author} {\bibfnamefont
  {P.}~\bibnamefont {Delsing}}, \bibinfo {author} {\bibfnamefont
  {J.}~\bibnamefont {Bylander}},\ and\ \bibinfo {author} {\bibfnamefont
  {A.}~\bibnamefont {Fadavi~Roudsari}},\ }\bibfield  {title} {\bibinfo {title}
  {Simplified {Josephson}-junction fabrication process for reproducibly
  high-performance superconducting qubits},\ }\href
  {https://doi.org/10.1063/5.0037093} {\bibfield  {journal} {\bibinfo
  {journal} {Appl. Phys. Lett.}\ }\textbf {\bibinfo {volume} {118}},\ \bibinfo
  {pages} {064002} (\bibinfo {year} {2021})}\BibitemShut {NoStop}%
\bibitem [{\citenamefont {Kim}\ \emph {et~al.}(2022)\citenamefont {Kim},
  \citenamefont {J\"unger}, \citenamefont {Morvan}, \citenamefont {Barnard},
  \citenamefont {Livingston}, \citenamefont {Alto\'e}, \citenamefont {Kim},
  \citenamefont {Song}, \citenamefont {Chen}, \citenamefont {Kreikebaum},
  \citenamefont {Ogletree}, \citenamefont {Santiago},\ and\ \citenamefont
  {Siddiqi}}]{Siddiqi2022}%
  \BibitemOpen
  \bibfield  {author} {\bibinfo {author} {\bibfnamefont {H.}~\bibnamefont
  {Kim}}, \bibinfo {author} {\bibfnamefont {C.}~\bibnamefont {J\"unger}},
  \bibinfo {author} {\bibfnamefont {A.}~\bibnamefont {Morvan}}, \bibinfo
  {author} {\bibfnamefont {E.~S.}\ \bibnamefont {Barnard}}, \bibinfo {author}
  {\bibfnamefont {W.~P.}\ \bibnamefont {Livingston}}, \bibinfo {author}
  {\bibfnamefont {M.~V.~P.}\ \bibnamefont {Alto\'e}}, \bibinfo {author}
  {\bibfnamefont {Y.}~\bibnamefont {Kim}}, \bibinfo {author} {\bibfnamefont
  {C.}~\bibnamefont {Song}}, \bibinfo {author} {\bibfnamefont {L.}~\bibnamefont
  {Chen}}, \bibinfo {author} {\bibfnamefont {J.~M.}\ \bibnamefont
  {Kreikebaum}}, \bibinfo {author} {\bibfnamefont {D.~F.}\ \bibnamefont
  {Ogletree}}, \bibinfo {author} {\bibfnamefont {D.~I.}\ \bibnamefont
  {Santiago}},\ and\ \bibinfo {author} {\bibfnamefont {I.}~\bibnamefont
  {Siddiqi}},\ }\bibfield  {title} {\bibinfo {title} {Effects of
  laser-annealing on fixed-frequency superconducting qubits},\ }\href
  {https://doi.org/10.1063/5.0102092} {\bibfield  {journal} {\bibinfo
  {journal} {Appl. Phys. Lett.}\ }\textbf {\bibinfo {volume} {121}},\ \bibinfo
  {pages} {142601} (\bibinfo {year} {2022})}\BibitemShut {NoStop}%
\bibitem [{\citenamefont {Bilmes}\ \emph {et~al.}(2021)\citenamefont {Bilmes},
  \citenamefont {H\"{a}ndel}, \citenamefont {Volosheniuk}, \citenamefont
  {Ustinov},\ and\ \citenamefont {Lisenfeld}}]{Bilmes2021}%
  \BibitemOpen
  \bibfield  {author} {\bibinfo {author} {\bibfnamefont {A.}~\bibnamefont
  {Bilmes}}, \bibinfo {author} {\bibfnamefont {A.~K.}\ \bibnamefont
  {H\"{a}ndel}}, \bibinfo {author} {\bibfnamefont {S.}~\bibnamefont
  {Volosheniuk}}, \bibinfo {author} {\bibfnamefont {A.~V.}\ \bibnamefont
  {Ustinov}},\ and\ \bibinfo {author} {\bibfnamefont {J.}~\bibnamefont
  {Lisenfeld}},\ }\bibfield  {title} {\bibinfo {title} {In-situ bandaged
  josephson junctions for superconducting quantum processors},\ }\href
  {https://doi.org/10.1088/1361-6668/ac2a6d} {\bibfield  {journal} {\bibinfo
  {journal} {Supercond. Sci. Technol.}\ }\textbf {\bibinfo {volume} {34}},\
  \bibinfo {pages} {125011} (\bibinfo {year} {2021})}\BibitemShut {NoStop}%
\bibitem [{\citenamefont {Premkumar}\ \emph {et~al.}(2021)\citenamefont
  {Premkumar}, \citenamefont {Weiland}, \citenamefont {Hwang}, \citenamefont
  {J{\"a}ck}, \citenamefont {Place}, \citenamefont {Waluyo}, \citenamefont
  {Hunt}, \citenamefont {Bisogni}, \citenamefont {Pelliciari}, \citenamefont
  {Barbour}, \citenamefont {Miller}, \citenamefont {Russo}, \citenamefont
  {Camino}, \citenamefont {Kisslinger}, \citenamefont {Tong}, \citenamefont
  {Hybertsen}, \citenamefont {Houck},\ and\ \citenamefont
  {Jarrige}}]{Premkumar2021}%
  \BibitemOpen
  \bibfield  {author} {\bibinfo {author} {\bibfnamefont {A.}~\bibnamefont
  {Premkumar}}, \bibinfo {author} {\bibfnamefont {C.}~\bibnamefont {Weiland}},
  \bibinfo {author} {\bibfnamefont {S.}~\bibnamefont {Hwang}}, \bibinfo
  {author} {\bibfnamefont {B.}~\bibnamefont {J{\"a}ck}}, \bibinfo {author}
  {\bibfnamefont {A.~P.~M.}\ \bibnamefont {Place}}, \bibinfo {author}
  {\bibfnamefont {I.}~\bibnamefont {Waluyo}}, \bibinfo {author} {\bibfnamefont
  {A.}~\bibnamefont {Hunt}}, \bibinfo {author} {\bibfnamefont {V.}~\bibnamefont
  {Bisogni}}, \bibinfo {author} {\bibfnamefont {J.}~\bibnamefont {Pelliciari}},
  \bibinfo {author} {\bibfnamefont {A.}~\bibnamefont {Barbour}}, \bibinfo
  {author} {\bibfnamefont {M.~S.}\ \bibnamefont {Miller}}, \bibinfo {author}
  {\bibfnamefont {P.}~\bibnamefont {Russo}}, \bibinfo {author} {\bibfnamefont
  {F.}~\bibnamefont {Camino}}, \bibinfo {author} {\bibfnamefont
  {K.}~\bibnamefont {Kisslinger}}, \bibinfo {author} {\bibfnamefont
  {X.}~\bibnamefont {Tong}}, \bibinfo {author} {\bibfnamefont {M.~S.}\
  \bibnamefont {Hybertsen}}, \bibinfo {author} {\bibfnamefont {A.~A.}\
  \bibnamefont {Houck}},\ and\ \bibinfo {author} {\bibfnamefont
  {I.}~\bibnamefont {Jarrige}},\ }\bibfield  {title} {\bibinfo {title}
  {Microscopic relaxation channels in materials for superconducting qubits},\
  }\href {https://doi.org/10.1038/s43246-021-00174-7} {\bibfield  {journal}
  {\bibinfo  {journal} {Commun. Mater.}\ }\textbf {\bibinfo {volume} {2}},\
  \bibinfo {pages} {72} (\bibinfo {year} {2021})}\BibitemShut {NoStop}%
\bibitem [{\citenamefont {Place}\ \emph {et~al.}(2021)\citenamefont {Place},
  \citenamefont {Rodgers}, \citenamefont {Mundada}, \citenamefont {Smitham},
  \citenamefont {Fitzpatrick}, \citenamefont {Leng}, \citenamefont {Premkumar},
  \citenamefont {Bryon}, \citenamefont {Vrajitoarea}, \citenamefont {Sussman},
  \citenamefont {Cheng}, \citenamefont {Madhavan}, \citenamefont {Babla},
  \citenamefont {Le}, \citenamefont {Gang}, \citenamefont {J{\"a}ck},
  \citenamefont {Gyenis}, \citenamefont {Yao}, \citenamefont {Cava},
  \citenamefont {de~Leon},\ and\ \citenamefont {Houck}}]{APMPlace2021}%
  \BibitemOpen
  \bibfield  {author} {\bibinfo {author} {\bibfnamefont {A.~P.~M.}\
  \bibnamefont {Place}}, \bibinfo {author} {\bibfnamefont {L.~V.~H.}\
  \bibnamefont {Rodgers}}, \bibinfo {author} {\bibfnamefont {P.}~\bibnamefont
  {Mundada}}, \bibinfo {author} {\bibfnamefont {B.~M.}\ \bibnamefont
  {Smitham}}, \bibinfo {author} {\bibfnamefont {M.}~\bibnamefont
  {Fitzpatrick}}, \bibinfo {author} {\bibfnamefont {Z.}~\bibnamefont {Leng}},
  \bibinfo {author} {\bibfnamefont {A.}~\bibnamefont {Premkumar}}, \bibinfo
  {author} {\bibfnamefont {J.}~\bibnamefont {Bryon}}, \bibinfo {author}
  {\bibfnamefont {A.}~\bibnamefont {Vrajitoarea}}, \bibinfo {author}
  {\bibfnamefont {S.}~\bibnamefont {Sussman}}, \bibinfo {author} {\bibfnamefont
  {G.}~\bibnamefont {Cheng}}, \bibinfo {author} {\bibfnamefont
  {T.}~\bibnamefont {Madhavan}}, \bibinfo {author} {\bibfnamefont {H.~K.}\
  \bibnamefont {Babla}}, \bibinfo {author} {\bibfnamefont {X.~H.}\ \bibnamefont
  {Le}}, \bibinfo {author} {\bibfnamefont {Y.}~\bibnamefont {Gang}}, \bibinfo
  {author} {\bibfnamefont {B.}~\bibnamefont {J{\"a}ck}}, \bibinfo {author}
  {\bibfnamefont {A.}~\bibnamefont {Gyenis}}, \bibinfo {author} {\bibfnamefont
  {N.}~\bibnamefont {Yao}}, \bibinfo {author} {\bibfnamefont {R.~J.}\
  \bibnamefont {Cava}}, \bibinfo {author} {\bibfnamefont {N.~P.}\ \bibnamefont
  {de~Leon}},\ and\ \bibinfo {author} {\bibfnamefont {A.~A.}\ \bibnamefont
  {Houck}},\ }\bibfield  {title} {\bibinfo {title} {New material platform for
  superconducting transmon qubits with coherence times exceeding 0.3
  milliseconds},\ }\href {https://doi.org/10.1038/s41467-021-22030-5}
  {\bibfield  {journal} {\bibinfo  {journal} {Nature Commun.}\ }\textbf
  {\bibinfo {volume} {12}},\ \bibinfo {pages} {1779} (\bibinfo {year}
  {2021})}\BibitemShut {NoStop}%
\bibitem [{\citenamefont {Chang}\ \emph {et~al.}(2013)\citenamefont {Chang},
  \citenamefont {Vissers}, \citenamefont {C\'orcoles}, \citenamefont
  {Sandberg}, \citenamefont {Gao}, \citenamefont {Abraham}, \citenamefont
  {Chow}, \citenamefont {Gambetta}, \citenamefont {Rothwell}, \citenamefont
  {Keefe}, \citenamefont {Steffen},\ and\ \citenamefont {Pappas}}]{Pappas2003}%
  \BibitemOpen
  \bibfield  {author} {\bibinfo {author} {\bibfnamefont {J.~B.}\ \bibnamefont
  {Chang}}, \bibinfo {author} {\bibfnamefont {M.~R.}\ \bibnamefont {Vissers}},
  \bibinfo {author} {\bibfnamefont {A.~D.}\ \bibnamefont {C\'orcoles}},
  \bibinfo {author} {\bibfnamefont {M.}~\bibnamefont {Sandberg}}, \bibinfo
  {author} {\bibfnamefont {J.}~\bibnamefont {Gao}}, \bibinfo {author}
  {\bibfnamefont {D.~W.}\ \bibnamefont {Abraham}}, \bibinfo {author}
  {\bibfnamefont {J.~M.}\ \bibnamefont {Chow}}, \bibinfo {author}
  {\bibfnamefont {J.~M.}\ \bibnamefont {Gambetta}}, \bibinfo {author}
  {\bibfnamefont {M.~B.}\ \bibnamefont {Rothwell}}, \bibinfo {author}
  {\bibfnamefont {G.~A.}\ \bibnamefont {Keefe}}, \bibinfo {author}
  {\bibfnamefont {M.}~\bibnamefont {Steffen}},\ and\ \bibinfo {author}
  {\bibfnamefont {D.~P.}\ \bibnamefont {Pappas}},\ }\bibfield  {title}
  {\bibinfo {title} {Improved superconducting qubit coherence using titanium
  nitride},\ }\href {https://doi.org/10.1063/1.4813269} {\bibfield  {journal}
  {\bibinfo  {journal} {Appl. Phys. Lett.}\ }\textbf {\bibinfo {volume}
  {103}},\ \bibinfo {pages} {012602} (\bibinfo {year} {2013})}\BibitemShut
  {NoStop}%
\bibitem [{\citenamefont {Crowley}\ \emph {et~al.}(2023)\citenamefont
  {Crowley}, \citenamefont {McLellan}, \citenamefont {Dutta}, \citenamefont
  {Shumiya}, \citenamefont {Place}, \citenamefont {Le}, \citenamefont {Gang},
  \citenamefont {Madhavan}, \citenamefont {Bland}, \citenamefont {Chang},
  \citenamefont {Khedkar}, \citenamefont {Feng}, \citenamefont {Umbarkar},
  \citenamefont {Gui}, \citenamefont {Rodgers}, \citenamefont {Jia},
  \citenamefont {Feldman}, \citenamefont {Lyon}, \citenamefont {Liu},
  \citenamefont {Cava}, \citenamefont {Houck},\ and\ \citenamefont
  {de~Leon}}]{Crowley2023}%
  \BibitemOpen
  \bibfield  {author} {\bibinfo {author} {\bibfnamefont {K.~D.}\ \bibnamefont
  {Crowley}}, \bibinfo {author} {\bibfnamefont {R.~A.}\ \bibnamefont
  {McLellan}}, \bibinfo {author} {\bibfnamefont {A.}~\bibnamefont {Dutta}},
  \bibinfo {author} {\bibfnamefont {N.}~\bibnamefont {Shumiya}}, \bibinfo
  {author} {\bibfnamefont {A.~P.~M.}\ \bibnamefont {Place}}, \bibinfo {author}
  {\bibfnamefont {X.~H.}\ \bibnamefont {Le}}, \bibinfo {author} {\bibfnamefont
  {Y.}~\bibnamefont {Gang}}, \bibinfo {author} {\bibfnamefont {T.}~\bibnamefont
  {Madhavan}}, \bibinfo {author} {\bibfnamefont {M.~P.}\ \bibnamefont {Bland}},
  \bibinfo {author} {\bibfnamefont {R.}~\bibnamefont {Chang}}, \bibinfo
  {author} {\bibfnamefont {N.}~\bibnamefont {Khedkar}}, \bibinfo {author}
  {\bibfnamefont {Y.~C.}\ \bibnamefont {Feng}}, \bibinfo {author}
  {\bibfnamefont {E.~A.}\ \bibnamefont {Umbarkar}}, \bibinfo {author}
  {\bibfnamefont {X.}~\bibnamefont {Gui}}, \bibinfo {author} {\bibfnamefont
  {L.~V.~H.}\ \bibnamefont {Rodgers}}, \bibinfo {author} {\bibfnamefont
  {Y.}~\bibnamefont {Jia}}, \bibinfo {author} {\bibfnamefont {M.~M.}\
  \bibnamefont {Feldman}}, \bibinfo {author} {\bibfnamefont {S.~A.}\
  \bibnamefont {Lyon}}, \bibinfo {author} {\bibfnamefont {M.}~\bibnamefont
  {Liu}}, \bibinfo {author} {\bibfnamefont {R.~J.}\ \bibnamefont {Cava}},
  \bibinfo {author} {\bibfnamefont {A.~A.}\ \bibnamefont {Houck}},\ and\
  \bibinfo {author} {\bibfnamefont {N.~P.}\ \bibnamefont {de~Leon}},\
  }\bibfield  {title} {\bibinfo {title} {Disentangling losses in tantalum
  superconducting circuits},\ }\href
  {https://doi.org/10.1103/PhysRevX.13.041005} {\bibfield  {journal} {\bibinfo
  {journal} {Phys. Rev. X}\ }\textbf {\bibinfo {volume} {13}},\ \bibinfo
  {pages} {041005} (\bibinfo {year} {2023})}\BibitemShut {NoStop}%
\bibitem [{\citenamefont {McLellan}\ \emph {et~al.}(2023)\citenamefont
  {McLellan}, \citenamefont {Dutta}, \citenamefont {Zhou}, \citenamefont {Jia},
  \citenamefont {Weiland}, \citenamefont {Gui}, \citenamefont {Place},
  \citenamefont {Crowley}, \citenamefont {Le}, \citenamefont {Madhavan},
  \citenamefont {Gang}, \citenamefont {Baker}, \citenamefont {Head},
  \citenamefont {Waluyo}, \citenamefont {Li}, \citenamefont {Kisslinger},
  \citenamefont {Hunt}, \citenamefont {Jarrige}, \citenamefont {Lyon},
  \citenamefont {Barbour}, \citenamefont {Cava}, \citenamefont {Houck},
  \citenamefont {Hulbert}, \citenamefont {Liu}, \citenamefont {Walter},\ and\
  \citenamefont {de~Leon}}]{McLellan2023}%
  \BibitemOpen
  \bibfield  {author} {\bibinfo {author} {\bibfnamefont {R.~A.}\ \bibnamefont
  {McLellan}}, \bibinfo {author} {\bibfnamefont {A.}~\bibnamefont {Dutta}},
  \bibinfo {author} {\bibfnamefont {C.}~\bibnamefont {Zhou}}, \bibinfo {author}
  {\bibfnamefont {Y.}~\bibnamefont {Jia}}, \bibinfo {author} {\bibfnamefont
  {C.}~\bibnamefont {Weiland}}, \bibinfo {author} {\bibfnamefont
  {X.}~\bibnamefont {Gui}}, \bibinfo {author} {\bibfnamefont {A.~P.~M.}\
  \bibnamefont {Place}}, \bibinfo {author} {\bibfnamefont {K.~D.}\ \bibnamefont
  {Crowley}}, \bibinfo {author} {\bibfnamefont {X.~H.}\ \bibnamefont {Le}},
  \bibinfo {author} {\bibfnamefont {T.}~\bibnamefont {Madhavan}}, \bibinfo
  {author} {\bibfnamefont {Y.}~\bibnamefont {Gang}}, \bibinfo {author}
  {\bibfnamefont {L.}~\bibnamefont {Baker}}, \bibinfo {author} {\bibfnamefont
  {A.~R.}\ \bibnamefont {Head}}, \bibinfo {author} {\bibfnamefont
  {I.}~\bibnamefont {Waluyo}}, \bibinfo {author} {\bibfnamefont
  {R.}~\bibnamefont {Li}}, \bibinfo {author} {\bibfnamefont {K.}~\bibnamefont
  {Kisslinger}}, \bibinfo {author} {\bibfnamefont {A.}~\bibnamefont {Hunt}},
  \bibinfo {author} {\bibfnamefont {I.}~\bibnamefont {Jarrige}}, \bibinfo
  {author} {\bibfnamefont {S.~A.}\ \bibnamefont {Lyon}}, \bibinfo {author}
  {\bibfnamefont {A.~M.}\ \bibnamefont {Barbour}}, \bibinfo {author}
  {\bibfnamefont {R.~J.}\ \bibnamefont {Cava}}, \bibinfo {author}
  {\bibfnamefont {A.~A.}\ \bibnamefont {Houck}}, \bibinfo {author}
  {\bibfnamefont {S.~L.}\ \bibnamefont {Hulbert}}, \bibinfo {author}
  {\bibfnamefont {M.}~\bibnamefont {Liu}}, \bibinfo {author} {\bibfnamefont
  {A.~L.}\ \bibnamefont {Walter}},\ and\ \bibinfo {author} {\bibfnamefont
  {N.~P.}\ \bibnamefont {de~Leon}},\ }\bibfield  {title} {\bibinfo {title}
  {Chemical profiles of the oxides on tantalum in state of the art
  superconducting circuits},\ }\href
  {https://doi.org/https://doi.org/10.1002/advs.202300921} {\bibfield
  {journal} {\bibinfo  {journal} {Adv. Sci.}\ }\textbf {\bibinfo {volume}
  {10}},\ \bibinfo {pages} {2300921} (\bibinfo {year} {2023})}\BibitemShut
  {NoStop}%
\bibitem [{\citenamefont {Williams}\ \emph {et~al.}(2003)\citenamefont
  {Williams}, \citenamefont {Gupta},\ and\ \citenamefont
  {Wasilik}}]{Williams2003}%
  \BibitemOpen
  \bibfield  {author} {\bibinfo {author} {\bibfnamefont {K.}~\bibnamefont
  {Williams}}, \bibinfo {author} {\bibfnamefont {K.}~\bibnamefont {Gupta}},\
  and\ \bibinfo {author} {\bibfnamefont {M.}~\bibnamefont {Wasilik}},\
  }\bibfield  {title} {\bibinfo {title} {Etch rates for micromachining
  processing-part ii},\ }\href {https://doi.org/10.1109/JMEMS.2003.820936}
  {\bibfield  {journal} {\bibinfo  {journal} {J. Microelectromech. Syst.}\
  }\textbf {\bibinfo {volume} {12}},\ \bibinfo {pages} {761} (\bibinfo {year}
  {2003})}\BibitemShut {NoStop}%
\bibitem [{\citenamefont {Koch}\ \emph {et~al.}(2007)\citenamefont {Koch},
  \citenamefont {Yu}, \citenamefont {Gambetta}, \citenamefont {Houck},
  \citenamefont {Schuster}, \citenamefont {Majer}, \citenamefont {Blais},
  \citenamefont {Devoret}, \citenamefont {Girvin},\ and\ \citenamefont
  {Schoelkopf}}]{Koch2007}%
  \BibitemOpen
  \bibfield  {author} {\bibinfo {author} {\bibfnamefont {J.}~\bibnamefont
  {Koch}}, \bibinfo {author} {\bibfnamefont {T.~M.}\ \bibnamefont {Yu}},
  \bibinfo {author} {\bibfnamefont {J.}~\bibnamefont {Gambetta}}, \bibinfo
  {author} {\bibfnamefont {A.~A.}\ \bibnamefont {Houck}}, \bibinfo {author}
  {\bibfnamefont {D.~I.}\ \bibnamefont {Schuster}}, \bibinfo {author}
  {\bibfnamefont {J.}~\bibnamefont {Majer}}, \bibinfo {author} {\bibfnamefont
  {A.}~\bibnamefont {Blais}}, \bibinfo {author} {\bibfnamefont {M.~H.}\
  \bibnamefont {Devoret}}, \bibinfo {author} {\bibfnamefont {S.~M.}\
  \bibnamefont {Girvin}},\ and\ \bibinfo {author} {\bibfnamefont {R.~J.}\
  \bibnamefont {Schoelkopf}},\ }\bibfield  {title} {\bibinfo {title}
  {Charge-insensitive qubit design derived from the {C}ooper pair box},\ }\href
  {https://doi.org/10.1103/PhysRevA.76.042319} {\bibfield  {journal} {\bibinfo
  {journal} {Phys. Rev. A}\ }\textbf {\bibinfo {volume} {76}},\ \bibinfo
  {pages} {042319} (\bibinfo {year} {2007})}\BibitemShut {NoStop}%
\bibitem [{\citenamefont {Kulik}\ and\ \citenamefont
  {Omel'yanchuk}(1975)}]{KO-1}%
  \BibitemOpen
  \bibfield  {author} {\bibinfo {author} {\bibfnamefont {I.~O.}\ \bibnamefont
  {Kulik}}\ and\ \bibinfo {author} {\bibfnamefont {A.~N.}\ \bibnamefont
  {Omel'yanchuk}},\ }\bibfield  {title} {\bibinfo {title} {Contribution to the
  microscopic theory of the {J}osephson effect in superconducting bridges},\
  }\href {https://www.osti.gov/biblio/4209268} {\bibfield  {journal} {\bibinfo
  {journal} {JETP Lett.}\ }\textbf {\bibinfo {volume} {21}},\ \bibinfo {pages}
  {96} (\bibinfo {year} {1975})}\BibitemShut {NoStop}%
\bibitem [{\citenamefont {Likharev}(1979)}]{Likharev1979}%
  \BibitemOpen
  \bibfield  {author} {\bibinfo {author} {\bibfnamefont {K.~K.}\ \bibnamefont
  {Likharev}},\ }\bibfield  {title} {\bibinfo {title} {Superconducting weak
  links},\ }\href {https://doi.org/10.1103/RevModPhys.51.101} {\bibfield
  {journal} {\bibinfo  {journal} {Rev. Mod. Phys.}\ }\textbf {\bibinfo {volume}
  {51}},\ \bibinfo {pages} {101} (\bibinfo {year} {1979})}\BibitemShut
  {NoStop}%
\bibitem [{\citenamefont {Golubov}\ \emph {et~al.}(2004)\citenamefont
  {Golubov}, \citenamefont {Kupriyanov},\ and\ \citenamefont
  {Il'ichev}}]{Golubov2004}%
  \BibitemOpen
  \bibfield  {author} {\bibinfo {author} {\bibfnamefont {A.~A.}\ \bibnamefont
  {Golubov}}, \bibinfo {author} {\bibfnamefont {M.~Y.}\ \bibnamefont
  {Kupriyanov}},\ and\ \bibinfo {author} {\bibfnamefont {E.}~\bibnamefont
  {Il'ichev}},\ }\bibfield  {title} {\bibinfo {title} {The current-phase
  relation in {J}osephson junctions},\ }\href
  {https://doi.org/10.1103/RevModPhys.76.411} {\bibfield  {journal} {\bibinfo
  {journal} {Rev. Mod. Phys.}\ }\textbf {\bibinfo {volume} {76}},\ \bibinfo
  {pages} {411} (\bibinfo {year} {2004})}\BibitemShut {NoStop}%
\bibitem [{\citenamefont {Vijay}\ \emph {et~al.}(2010)\citenamefont {Vijay},
  \citenamefont {Levenson-Falk}, \citenamefont {Slichter},\ and\ \citenamefont
  {Siddiqi}}]{Vijay2010}%
  \BibitemOpen
  \bibfield  {author} {\bibinfo {author} {\bibfnamefont {R.}~\bibnamefont
  {Vijay}}, \bibinfo {author} {\bibfnamefont {E.~M.}\ \bibnamefont
  {Levenson-Falk}}, \bibinfo {author} {\bibfnamefont {D.~H.}\ \bibnamefont
  {Slichter}},\ and\ \bibinfo {author} {\bibfnamefont {I.}~\bibnamefont
  {Siddiqi}},\ }\bibfield  {title} {\bibinfo {title} {Approaching ideal weak
  link behavior with three dimensional aluminum nanobridges},\ }\href
  {https://doi.org/10.1063/1.3443716} {\bibfield  {journal} {\bibinfo
  {journal} {Appl. Phys. Lett.}\ }\textbf {\bibinfo {volume} {96}},\ \bibinfo
  {pages} {223112} (\bibinfo {year} {2010})}\BibitemShut {NoStop}%
\bibitem [{\citenamefont {Hioe}\ and\ \citenamefont
  {Montroll}(1975)}]{Hioe1975}%
  \BibitemOpen
  \bibfield  {author} {\bibinfo {author} {\bibfnamefont {F.~T.}\ \bibnamefont
  {Hioe}}\ and\ \bibinfo {author} {\bibfnamefont {E.~W.}\ \bibnamefont
  {Montroll}},\ }\bibfield  {title} {\bibinfo {title} {{Quantum theory of
  anharmonic oscillators. I. Energy levels of oscillators with positive quartic
  anharmonicity}},\ }\href {https://doi.org/10.1063/1.522747} {\bibfield
  {journal} {\bibinfo  {journal} {J. Math. Phys.}\ }\textbf {\bibinfo {volume}
  {16}},\ \bibinfo {pages} {1945} (\bibinfo {year} {1975})}\BibitemShut
  {NoStop}%
\bibitem [{\citenamefont {Vijay}\ \emph {et~al.}(2009)\citenamefont {Vijay},
  \citenamefont {Sau}, \citenamefont {Cohen},\ and\ \citenamefont
  {Siddiqi}}]{Vijay2009}%
  \BibitemOpen
  \bibfield  {author} {\bibinfo {author} {\bibfnamefont {R.}~\bibnamefont
  {Vijay}}, \bibinfo {author} {\bibfnamefont {J.~D.}\ \bibnamefont {Sau}},
  \bibinfo {author} {\bibfnamefont {M.~L.}\ \bibnamefont {Cohen}},\ and\
  \bibinfo {author} {\bibfnamefont {I.}~\bibnamefont {Siddiqi}},\ }\bibfield
  {title} {\bibinfo {title} {Optimizing anharmonicity in nanoscale weak link
  josephson junction oscillators},\ }\href
  {https://doi.org/10.1103/PhysRevLett.103.087003} {\bibfield  {journal}
  {\bibinfo  {journal} {Phys. Rev. Lett.}\ }\textbf {\bibinfo {volume} {103}},\
  \bibinfo {pages} {087003} (\bibinfo {year} {2009})}\BibitemShut {NoStop}%
\bibitem [{\citenamefont {Hasselbach}\ \emph {et~al.}(2002)\citenamefont
  {Hasselbach}, \citenamefont {Mailly},\ and\ \citenamefont
  {Kirtley}}]{Hasselbach2002}%
  \BibitemOpen
  \bibfield  {author} {\bibinfo {author} {\bibfnamefont {K.}~\bibnamefont
  {Hasselbach}}, \bibinfo {author} {\bibfnamefont {D.}~\bibnamefont {Mailly}},\
  and\ \bibinfo {author} {\bibfnamefont {J.~R.}\ \bibnamefont {Kirtley}},\
  }\bibfield  {title} {\bibinfo {title} {{Micro-superconducting quantum
  interference device characteristics}},\ }\href
  {https://doi.org/10.1063/1.1448864} {\bibfield  {journal} {\bibinfo
  {journal} {J. Appl. Phys.}\ }\textbf {\bibinfo {volume} {91}},\ \bibinfo
  {pages} {4432} (\bibinfo {year} {2002})}\BibitemShut {NoStop}%
\bibitem [{\citenamefont {Lindelof}(1981)}]{Lindelof1981}%
  \BibitemOpen
  \bibfield  {author} {\bibinfo {author} {\bibfnamefont {P.~E.}\ \bibnamefont
  {Lindelof}},\ }\bibfield  {title} {\bibinfo {title} {Superconducting
  microbridges exhibiting josephson properties},\ }\href
  {https://doi.org/10.1088/0034-4885/44/9/001} {\bibfield  {journal} {\bibinfo
  {journal} {Rep. Prog. Phys.}\ }\textbf {\bibinfo {volume} {44}},\ \bibinfo
  {pages} {949} (\bibinfo {year} {1981})}\BibitemShut {NoStop}%
\bibitem [{\citenamefont {Tinkham}(2004)}]{TinkhamBook2004}%
  \BibitemOpen
  \bibfield  {author} {\bibinfo {author} {\bibfnamefont {M.}~\bibnamefont
  {Tinkham}},\ }\href@noop {} {\emph {\bibinfo {title} {Introduction to
  Superconductivity}}},\ \bibinfo {edition} {2nd}\ ed.\ (\bibinfo  {publisher}
  {Dover Publications},\ \bibinfo {year} {2004})\BibitemShut {NoStop}%
\bibitem [{\citenamefont {Likharev}\ and\ \citenamefont
  {Yakobson}(1975)}]{Likharev1975c}%
  \BibitemOpen
  \bibfield  {author} {\bibinfo {author} {\bibfnamefont {K.~K.}\ \bibnamefont
  {Likharev}}\ and\ \bibinfo {author} {\bibfnamefont {L.~A.}\ \bibnamefont
  {Yakobson}},\ }\bibfield  {title} {\bibinfo {title} {Steady-state properties
  of superconducting bridges},\ }\href@noop {} {\bibfield  {journal} {\bibinfo
  {journal} {Sov. Phys.-Tekhn. Phys.}\ }\textbf {\bibinfo {volume} {20}},\
  \bibinfo {pages} {950} (\bibinfo {year} {1975})}\BibitemShut {NoStop}%
\bibitem [{\citenamefont {Sch\"ussler}\ and\ \citenamefont
  {K\"ummel}(1993)}]{Schussler1993}%
  \BibitemOpen
  \bibfield  {author} {\bibinfo {author} {\bibfnamefont {U.}~\bibnamefont
  {Sch\"ussler}}\ and\ \bibinfo {author} {\bibfnamefont {R.}~\bibnamefont
  {K\"ummel}},\ }\bibfield  {title} {\bibinfo {title} {Andreev scattering,
  josephson currents, and coupling energy in clean
  superconductor-semiconductor-superconductor junctions},\ }\href
  {https://doi.org/10.1103/PhysRevB.47.2754} {\bibfield  {journal} {\bibinfo
  {journal} {Phys. Rev. B}\ }\textbf {\bibinfo {volume} {47}},\ \bibinfo
  {pages} {2754} (\bibinfo {year} {1993})}\BibitemShut {NoStop}%
\bibitem [{\citenamefont {Werninghaus}\ \emph {et~al.}(2021)\citenamefont
  {Werninghaus}, \citenamefont {Egger}, \citenamefont {Roy}, \citenamefont
  {Machnes}, \citenamefont {Wilhelm},\ and\ \citenamefont
  {Filipp}}]{Werninghaus2021}%
  \BibitemOpen
  \bibfield  {author} {\bibinfo {author} {\bibfnamefont {M.}~\bibnamefont
  {Werninghaus}}, \bibinfo {author} {\bibfnamefont {D.~J.}\ \bibnamefont
  {Egger}}, \bibinfo {author} {\bibfnamefont {F.}~\bibnamefont {Roy}}, \bibinfo
  {author} {\bibfnamefont {S.}~\bibnamefont {Machnes}}, \bibinfo {author}
  {\bibfnamefont {F.~K.}\ \bibnamefont {Wilhelm}},\ and\ \bibinfo {author}
  {\bibfnamefont {S.}~\bibnamefont {Filipp}},\ }\bibfield  {title} {\bibinfo
  {title} {Leakage reduction in fast superconducting qubit gates via optimal
  control},\ }\href {https://doi.org/10.1038/s41534-020-00346-2} {\bibfield
  {journal} {\bibinfo  {journal} {npj Quantum Inf.}\ }\textbf {\bibinfo
  {volume} {7}},\ \bibinfo {pages} {14} (\bibinfo {year} {2021})}\BibitemShut
  {NoStop}%
\bibitem [{\citenamefont {Kjaergaard}\ \emph {et~al.}(2020)\citenamefont
  {Kjaergaard}, \citenamefont {Schwartz}, \citenamefont {Braum\"{u}ller},
  \citenamefont {Krantz}, \citenamefont {Wang}, \citenamefont {Gustavsson},\
  and\ \citenamefont {Oliver}}]{Kjaergaard2020}%
  \BibitemOpen
  \bibfield  {author} {\bibinfo {author} {\bibfnamefont {M.}~\bibnamefont
  {Kjaergaard}}, \bibinfo {author} {\bibfnamefont {M.~E.}\ \bibnamefont
  {Schwartz}}, \bibinfo {author} {\bibfnamefont {J.}~\bibnamefont
  {Braum\"{u}ller}}, \bibinfo {author} {\bibfnamefont {P.}~\bibnamefont
  {Krantz}}, \bibinfo {author} {\bibfnamefont {J.~I.-J.}\ \bibnamefont {Wang}},
  \bibinfo {author} {\bibfnamefont {S.}~\bibnamefont {Gustavsson}},\ and\
  \bibinfo {author} {\bibfnamefont {W.~D.}\ \bibnamefont {Oliver}},\ }\bibfield
   {title} {\bibinfo {title} {Superconducting qubits: Current state of play},\
  }\href {https://doi.org/10.1146/annurev-conmatphys-031119-050605} {\bibfield
  {journal} {\bibinfo  {journal} {Annu. Rev. Condens. Matter Phys.}\ }\textbf
  {\bibinfo {volume} {11}},\ \bibinfo {pages} {369} (\bibinfo {year}
  {2020})}\BibitemShut {NoStop}%
\bibitem [{\citenamefont {Rotzinger}\ \emph {et~al.}(2016)\citenamefont
  {Rotzinger}, \citenamefont {Skacel}, \citenamefont {Pfirrmann}, \citenamefont
  {Voss}, \citenamefont {M\"{u}nzberg}, \citenamefont {Probst}, \citenamefont
  {Bushev}, \citenamefont {Weides}, \citenamefont {Ustinov},\ and\
  \citenamefont {Mooij}}]{Rotzinger2017}%
  \BibitemOpen
  \bibfield  {author} {\bibinfo {author} {\bibfnamefont {H.}~\bibnamefont
  {Rotzinger}}, \bibinfo {author} {\bibfnamefont {S.~T.}\ \bibnamefont
  {Skacel}}, \bibinfo {author} {\bibfnamefont {M.}~\bibnamefont {Pfirrmann}},
  \bibinfo {author} {\bibfnamefont {J.~N.}\ \bibnamefont {Voss}}, \bibinfo
  {author} {\bibfnamefont {J.}~\bibnamefont {M\"{u}nzberg}}, \bibinfo {author}
  {\bibfnamefont {S.}~\bibnamefont {Probst}}, \bibinfo {author} {\bibfnamefont
  {P.}~\bibnamefont {Bushev}}, \bibinfo {author} {\bibfnamefont {M.~P.}\
  \bibnamefont {Weides}}, \bibinfo {author} {\bibfnamefont {A.~V.}\
  \bibnamefont {Ustinov}},\ and\ \bibinfo {author} {\bibfnamefont {J.~E.}\
  \bibnamefont {Mooij}},\ }\bibfield  {title} {\bibinfo {title}
  {Aluminium-oxide wires for superconducting high kinetic inductance
  circuits},\ }\href {https://doi.org/10.1088/0953-2048/30/2/025002} {\bibfield
   {journal} {\bibinfo  {journal} {Supercond. Sci. Technol.}\ }\textbf
  {\bibinfo {volume} {30}},\ \bibinfo {pages} {025002} (\bibinfo {year}
  {2016})}\BibitemShut {NoStop}%
\bibitem [{\citenamefont {Maleeva}\ \emph {et~al.}(2018)\citenamefont
  {Maleeva}, \citenamefont {Gr{\"u}nhaupt}, \citenamefont {Klein},
  \citenamefont {Levy-Bertrand}, \citenamefont {Dupre}, \citenamefont {Calvo},
  \citenamefont {Valenti}, \citenamefont {Winkel}, \citenamefont {Friedrich},
  \citenamefont {Wernsdorfer}, \citenamefont {Ustinov}, \citenamefont
  {Rotzinger}, \citenamefont {Monfardini}, \citenamefont {Fistul},\ and\
  \citenamefont {Pop}}]{Maleeva2018}%
  \BibitemOpen
  \bibfield  {author} {\bibinfo {author} {\bibfnamefont {N.}~\bibnamefont
  {Maleeva}}, \bibinfo {author} {\bibfnamefont {L.}~\bibnamefont
  {Gr{\"u}nhaupt}}, \bibinfo {author} {\bibfnamefont {T.}~\bibnamefont
  {Klein}}, \bibinfo {author} {\bibfnamefont {F.}~\bibnamefont
  {Levy-Bertrand}}, \bibinfo {author} {\bibfnamefont {O.}~\bibnamefont
  {Dupre}}, \bibinfo {author} {\bibfnamefont {M.}~\bibnamefont {Calvo}},
  \bibinfo {author} {\bibfnamefont {F.}~\bibnamefont {Valenti}}, \bibinfo
  {author} {\bibfnamefont {P.}~\bibnamefont {Winkel}}, \bibinfo {author}
  {\bibfnamefont {F.}~\bibnamefont {Friedrich}}, \bibinfo {author}
  {\bibfnamefont {W.}~\bibnamefont {Wernsdorfer}}, \bibinfo {author}
  {\bibfnamefont {A.~V.}\ \bibnamefont {Ustinov}}, \bibinfo {author}
  {\bibfnamefont {H.}~\bibnamefont {Rotzinger}}, \bibinfo {author}
  {\bibfnamefont {A.}~\bibnamefont {Monfardini}}, \bibinfo {author}
  {\bibfnamefont {M.~V.}\ \bibnamefont {Fistul}},\ and\ \bibinfo {author}
  {\bibfnamefont {I.~M.}\ \bibnamefont {Pop}},\ }\bibfield  {title} {\bibinfo
  {title} {Circuit quantum electrodynamics of granular aluminum resonators},\
  }\href {https://doi.org/10.1038/s41467-018-06386-9} {\bibfield  {journal}
  {\bibinfo  {journal} {Nature Commun.}\ }\textbf {\bibinfo {volume} {9}},\
  \bibinfo {pages} {3889} (\bibinfo {year} {2018})}\BibitemShut {NoStop}%
\bibitem [{\citenamefont {Marrache-Kikuchi}\ \emph {et~al.}(2008)\citenamefont
  {Marrache-Kikuchi}, \citenamefont {Aubin}, \citenamefont {Pourret},
  \citenamefont {Behnia}, \citenamefont {Lesueur}, \citenamefont {Berg\'e},\
  and\ \citenamefont {Dumoulin}}]{Dumoulin2008}%
  \BibitemOpen
  \bibfield  {author} {\bibinfo {author} {\bibfnamefont {C.~A.}\ \bibnamefont
  {Marrache-Kikuchi}}, \bibinfo {author} {\bibfnamefont {H.}~\bibnamefont
  {Aubin}}, \bibinfo {author} {\bibfnamefont {A.}~\bibnamefont {Pourret}},
  \bibinfo {author} {\bibfnamefont {K.}~\bibnamefont {Behnia}}, \bibinfo
  {author} {\bibfnamefont {J.}~\bibnamefont {Lesueur}}, \bibinfo {author}
  {\bibfnamefont {L.}~\bibnamefont {Berg\'e}},\ and\ \bibinfo {author}
  {\bibfnamefont {L.}~\bibnamefont {Dumoulin}},\ }\bibfield  {title} {\bibinfo
  {title} {Thickness-tuned superconductor-insulator transitions under magnetic
  field in $a$-{NbSi}},\ }\href {https://doi.org/10.1103/PhysRevB.78.144520}
  {\bibfield  {journal} {\bibinfo  {journal} {Phys. Rev. B}\ }\textbf {\bibinfo
  {volume} {78}},\ \bibinfo {pages} {144520} (\bibinfo {year}
  {2008})}\BibitemShut {NoStop}%
\bibitem [{\citenamefont {Crauste}\ \emph {et~al.}(2013)\citenamefont
  {Crauste}, \citenamefont {Gentils}, \citenamefont {Cou\"edo}, \citenamefont
  {Dolgorouky}, \citenamefont {Berg\'e}, \citenamefont {Collin}, \citenamefont
  {Marrache-Kikuchi},\ and\ \citenamefont {Dumoulin}}]{Dumoulin2013}%
  \BibitemOpen
  \bibfield  {author} {\bibinfo {author} {\bibfnamefont {O.}~\bibnamefont
  {Crauste}}, \bibinfo {author} {\bibfnamefont {A.}~\bibnamefont {Gentils}},
  \bibinfo {author} {\bibfnamefont {F.}~\bibnamefont {Cou\"edo}}, \bibinfo
  {author} {\bibfnamefont {Y.}~\bibnamefont {Dolgorouky}}, \bibinfo {author}
  {\bibfnamefont {L.}~\bibnamefont {Berg\'e}}, \bibinfo {author} {\bibfnamefont
  {S.}~\bibnamefont {Collin}}, \bibinfo {author} {\bibfnamefont {C.~A.}\
  \bibnamefont {Marrache-Kikuchi}},\ and\ \bibinfo {author} {\bibfnamefont
  {L.}~\bibnamefont {Dumoulin}},\ }\bibfield  {title} {\bibinfo {title} {Effect
  of annealing on the superconducting properties of
  $a$-{Nb}${}_{x}${Si}${}_{1\ensuremath{-}x}$ thin films},\ }\href
  {https://doi.org/10.1103/PhysRevB.87.144514} {\bibfield  {journal} {\bibinfo
  {journal} {Phys. Rev. B}\ }\textbf {\bibinfo {volume} {87}},\ \bibinfo
  {pages} {144514} (\bibinfo {year} {2013})}\BibitemShut {NoStop}%
\bibitem [{\citenamefont {Webster}\ \emph {et~al.}(2013)\citenamefont
  {Webster}, \citenamefont {Fenton}, \citenamefont {Hongisto}, \citenamefont
  {Giblin}, \citenamefont {Zorin},\ and\ \citenamefont
  {Warburton}}]{Webster2013}%
  \BibitemOpen
  \bibfield  {author} {\bibinfo {author} {\bibfnamefont {C.~H.}\ \bibnamefont
  {Webster}}, \bibinfo {author} {\bibfnamefont {J.~C.}\ \bibnamefont {Fenton}},
  \bibinfo {author} {\bibfnamefont {T.~T.}\ \bibnamefont {Hongisto}}, \bibinfo
  {author} {\bibfnamefont {S.~P.}\ \bibnamefont {Giblin}}, \bibinfo {author}
  {\bibfnamefont {A.~B.}\ \bibnamefont {Zorin}},\ and\ \bibinfo {author}
  {\bibfnamefont {P.~A.}\ \bibnamefont {Warburton}},\ }\bibfield  {title}
  {\bibinfo {title} {{NbSi} nanowire quantum phase-slip circuits: dc
  supercurrent blockade, microwave measurements, and thermal analysis},\ }\href
  {https://doi.org/10.1103/PhysRevB.87.144510} {\bibfield  {journal} {\bibinfo
  {journal} {Phys. Rev. B}\ }\textbf {\bibinfo {volume} {87}},\ \bibinfo
  {pages} {144510} (\bibinfo {year} {2013})}\BibitemShut {NoStop}%
\bibitem [{\citenamefont {Fornberg}(1988)}]{Fornberg1988}%
  \BibitemOpen
  \bibfield  {author} {\bibinfo {author} {\bibfnamefont {B.}~\bibnamefont
  {Fornberg}},\ }\bibfield  {title} {\bibinfo {title} {Generation of finite
  difference formulas on arbitrarily spaced grids},\ }\href
  {https://doi.org/10.1090/S0025-5718-1988-0935077-0} {\bibfield  {journal}
  {\bibinfo  {journal} {Math. Comp.}\ }\textbf {\bibinfo {volume} {51}},\
  \bibinfo {pages} {699} (\bibinfo {year} {1988})}\BibitemShut {NoStop}%
\bibitem [{\citenamefont {Virtanen}\ \emph {et~al.}(2020)\citenamefont
  {Virtanen}, \citenamefont {Gommers}, \citenamefont {Oliphant}, \citenamefont
  {Haberland}, \citenamefont {Reddy}, \citenamefont {Cournapeau}, \citenamefont
  {Burovski}, \citenamefont {Peterson}, \citenamefont {Weckesser},
  \citenamefont {Bright}, \citenamefont {{van der Walt}}, \citenamefont
  {Brett}, \citenamefont {Wilson}, \citenamefont {Millman}, \citenamefont
  {Mayorov}, \citenamefont {Nelson}, \citenamefont {Jones}, \citenamefont
  {Kern}, \citenamefont {Larson}, \citenamefont {Carey}, \citenamefont {Polat},
  \citenamefont {Feng}, \citenamefont {Moore}, \citenamefont {{VanderPlas}},
  \citenamefont {Laxalde}, \citenamefont {Perktold}, \citenamefont {Cimrman},
  \citenamefont {Henriksen}, \citenamefont {Quintero}, \citenamefont {Harris},
  \citenamefont {Archibald}, \citenamefont {Ribeiro}, \citenamefont
  {Pedregosa}, \citenamefont {{van Mulbregt}},\ and\ \citenamefont {{SciPy 1.0
  Contributors}}}]{2020SciPy-NMeth}%
  \BibitemOpen
  \bibfield  {author} {\bibinfo {author} {\bibfnamefont {P.}~\bibnamefont
  {Virtanen}}, \bibinfo {author} {\bibfnamefont {R.}~\bibnamefont {Gommers}},
  \bibinfo {author} {\bibfnamefont {T.~E.}\ \bibnamefont {Oliphant}}, \bibinfo
  {author} {\bibfnamefont {M.}~\bibnamefont {Haberland}}, \bibinfo {author}
  {\bibfnamefont {T.}~\bibnamefont {Reddy}}, \bibinfo {author} {\bibfnamefont
  {D.}~\bibnamefont {Cournapeau}}, \bibinfo {author} {\bibfnamefont
  {E.}~\bibnamefont {Burovski}}, \bibinfo {author} {\bibfnamefont
  {P.}~\bibnamefont {Peterson}}, \bibinfo {author} {\bibfnamefont
  {W.}~\bibnamefont {Weckesser}}, \bibinfo {author} {\bibfnamefont
  {J.}~\bibnamefont {Bright}}, \bibinfo {author} {\bibfnamefont {S.~J.}\
  \bibnamefont {{van der Walt}}}, \bibinfo {author} {\bibfnamefont
  {M.}~\bibnamefont {Brett}}, \bibinfo {author} {\bibfnamefont
  {J.}~\bibnamefont {Wilson}}, \bibinfo {author} {\bibfnamefont {K.~J.}\
  \bibnamefont {Millman}}, \bibinfo {author} {\bibfnamefont {N.}~\bibnamefont
  {Mayorov}}, \bibinfo {author} {\bibfnamefont {A.~R.~J.}\ \bibnamefont
  {Nelson}}, \bibinfo {author} {\bibfnamefont {E.}~\bibnamefont {Jones}},
  \bibinfo {author} {\bibfnamefont {R.}~\bibnamefont {Kern}}, \bibinfo {author}
  {\bibfnamefont {E.}~\bibnamefont {Larson}}, \bibinfo {author} {\bibfnamefont
  {C.~J.}\ \bibnamefont {Carey}}, \bibinfo {author} {\bibfnamefont
  {{\.I}.}~\bibnamefont {Polat}}, \bibinfo {author} {\bibfnamefont
  {Y.}~\bibnamefont {Feng}}, \bibinfo {author} {\bibfnamefont {E.~W.}\
  \bibnamefont {Moore}}, \bibinfo {author} {\bibfnamefont {J.}~\bibnamefont
  {{VanderPlas}}}, \bibinfo {author} {\bibfnamefont {D.}~\bibnamefont
  {Laxalde}}, \bibinfo {author} {\bibfnamefont {J.}~\bibnamefont {Perktold}},
  \bibinfo {author} {\bibfnamefont {R.}~\bibnamefont {Cimrman}}, \bibinfo
  {author} {\bibfnamefont {I.}~\bibnamefont {Henriksen}}, \bibinfo {author}
  {\bibfnamefont {E.~A.}\ \bibnamefont {Quintero}}, \bibinfo {author}
  {\bibfnamefont {C.~R.}\ \bibnamefont {Harris}}, \bibinfo {author}
  {\bibfnamefont {A.~M.}\ \bibnamefont {Archibald}}, \bibinfo {author}
  {\bibfnamefont {A.~H.}\ \bibnamefont {Ribeiro}}, \bibinfo {author}
  {\bibfnamefont {F.}~\bibnamefont {Pedregosa}}, \bibinfo {author}
  {\bibfnamefont {P.}~\bibnamefont {{van Mulbregt}}},\ and\ \bibinfo {author}
  {\bibnamefont {{SciPy 1.0 Contributors}}},\ }\bibfield  {title} {\bibinfo
  {title} {{{SciPy} 1.0: Fundamental Algorithms for Scientific Computing in
  Python}},\ }\href {https://doi.org/10.1038/s41592-019-0686-2} {\bibfield
  {journal} {\bibinfo  {journal} {Nature Methods}\ }\textbf {\bibinfo {volume}
  {17}},\ \bibinfo {pages} {261} (\bibinfo {year} {2020})}\BibitemShut
  {NoStop}%
\end{thebibliography}

%

\end{document}